\documentclass[twocolumn]{aastex6}
\usepackage{graphicx}
\usepackage{amssymb,amsfonts,amsmath,amstext,amsgen,amsopn,amsxtra,indentfirst,times,subfigure,natbib}
\begin{document}
\title{Revisiting the Phase Curves of WASP-43b: Confronting Reanalyzed Spitzer Data \\ with Cloudy Atmospheres}
\author{Jo\~{a}o M. Mendon\c{c}a\altaffilmark{1,2,3}}
\author{Matej Malik\altaffilmark{3}}
\author{Brice-Olivier Demory\altaffilmark{3}}
\author{Kevin Heng\altaffilmark{3}}
\altaffiltext{1}{Astrophysics and Atmospheric Physics, National Space Institute, Technical University of Denmark, Elektrovej, 2800, Kgs. Lyngby, Denmark}
\altaffiltext{2}{University of Copenhagen, Centre for Star and Planet Formation, Niels Bohr Institute and Natural History Museum, DK-1350, Copenhagen, Denmark}
\altaffiltext{3}{University of Bern, Center for Space and Habitability, Gesellschaftsstrasse 6, CH-3012, Bern, Switzerland.  Emails: joao.mendonca@space.dtu.dk, matej.malik@csh.unibe.ch, brice.demory@csh.unibe.ch, kevin.heng@csh.unibe.ch}

\begin{abstract}
Recently acquired \textit{Hubble} and \textit{Spitzer} phase curves of the short-period hot Jupiter WASP-43b make it an ideal target for confronting theory with data.  On the observational front, we re-analyze the 3.6 and 4.5 $\mu$m \textit{Spitzer} phase curves and demonstrate that our improved analysis better removes residual red noise due to intra-pixel sensitivity, which leads to greater fluxes emanating from the nightside of WASP-43b, thus reducing the tension between theory and data.  On the theoretical front, we construct cloudfree and cloudy atmospheres of WASP-43b using our Global Circulation Model (GCM), \texttt{THOR}, which solves the non-hydrostatic Euler equations (compared to GCMs that typically solve the hydrostatic primitive equations).  The cloudfree atmosphere produces a reasonable fit to the dayside emission spectrum. The multi-phase emission spectra constrain the cloud deck to be confined to the nightside and have a finite cloud-top pressure.  The multi-wavelength phase curves are naturally consistent with our cloudy atmospheres, except for the 4.5 $\mu$m phase curve, which requires the presence of enhanced carbon dioxide in the atmosphere of WASP-43b.  Multi-phase emission spectra at higher spectral resolution, as may be obtained using the \textit{James Webb Space Telescope}, and a reflected-light phase curve at visible wavelengths would further constrain the properties of clouds in WASP-43b.
\end{abstract}

\section{Introduction}
\label{sec:intro}

\subsection{Background}

The hot Jupiter WASP-43b, which is about twice as massive and the same radius as Jupiter, orbits its K7 star in just 19.2 hours \citep{2012Gillon}, making it a prime target for phase curve observations using the \textit{Hubble Space Telescope (HST)} \citep{2014Stevenson}.  These multi-wavelength phase curves probe WASP-43b across longitude and depth (or pressure), providing a two-dimensional view of its atmosphere.  Other inferred properties of WASP-43b's atmosphere include a low dayside-nightside energy redistribution efficiency, the lack of a temperature inversion on its dayside and constraints on its water abundance \citep{2012Gillon,2013Wang,2014Blecic,2014Kreidberg,2014Stevenson}.

\subsection{Observational motivation}

To add to the set of near-infrared phase curves published by \citet{2014Stevenson}, two 3.6 and one 4.5 $\mu$m {\it Spitzer} phase curves were recently published by \citet{2017Stevenson}.  The authors report large differences between the two 3.6 $\mu$m phase-curve amplitudes and a deep signal (200--300 ppm) present in the second phase curve at orbital phase $\sim$0.6 that they choose to dismiss. \citet{2017Stevenson} further state that instrumental systematics affected the first 3.6 $\mu$m phase-curve and thus requested re-observations in the same channel. Because of the long-duration of these observations, the visits were split into three (for the first phase-curve) and two (for the second one) Astronomical Observation Requests (AOR), which were not adjacent in time. \citet{2017Stevenson} state that the origin of the increased correlated noise was due to the fact that the star landed on different areas of the pixel for each AOR, enhancing the impact of the well-documented intra-pixel sensitivity \citep[e.g.,][]{2012Ingalls} on the photometry. These facts motivated us to perform our own, independent analysis of the \textit{Spitzer} data.  

Thus, part of the present study is devoted to presenting a new re-analysis of the \textit{Spitzer} phase curves of WASP-43b.

\begin{figure*}
\begin{center}
\includegraphics[width=\columnwidth]{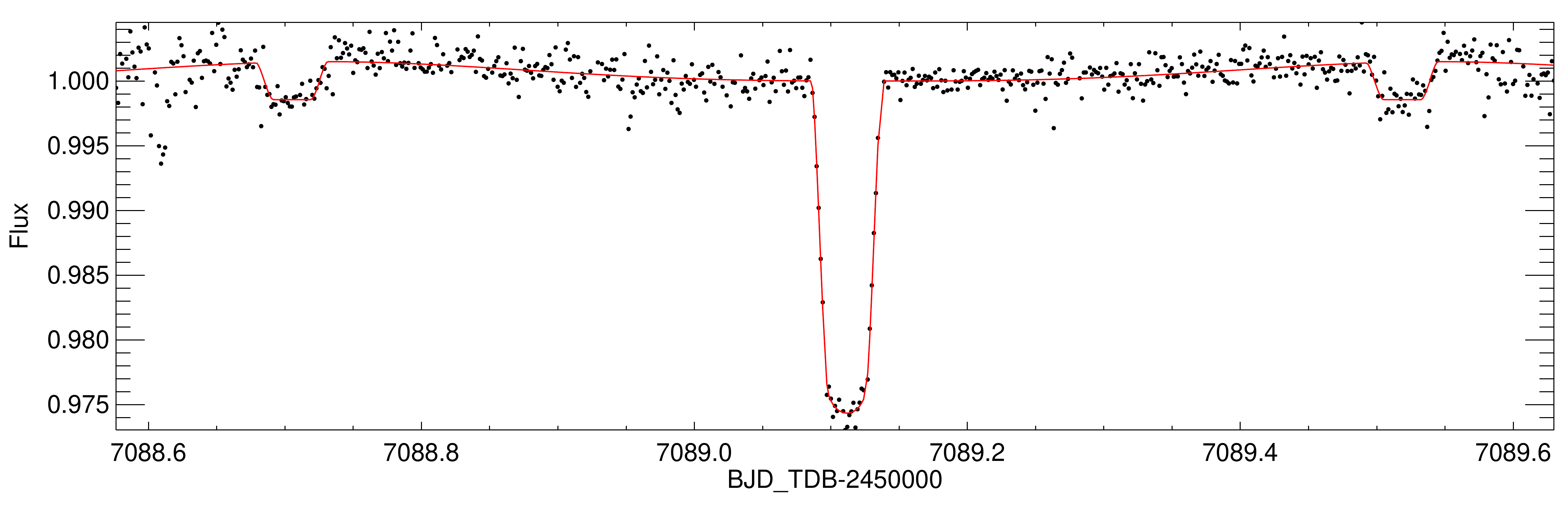}
\includegraphics[width=\columnwidth]{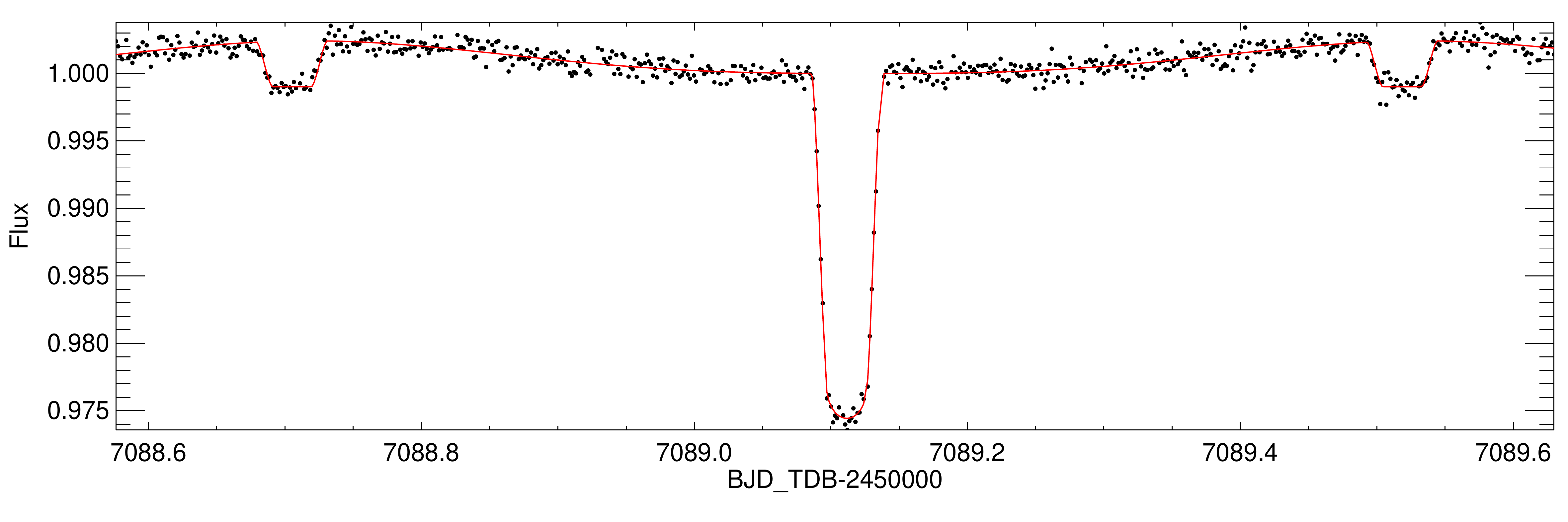}
\includegraphics[width=\columnwidth]{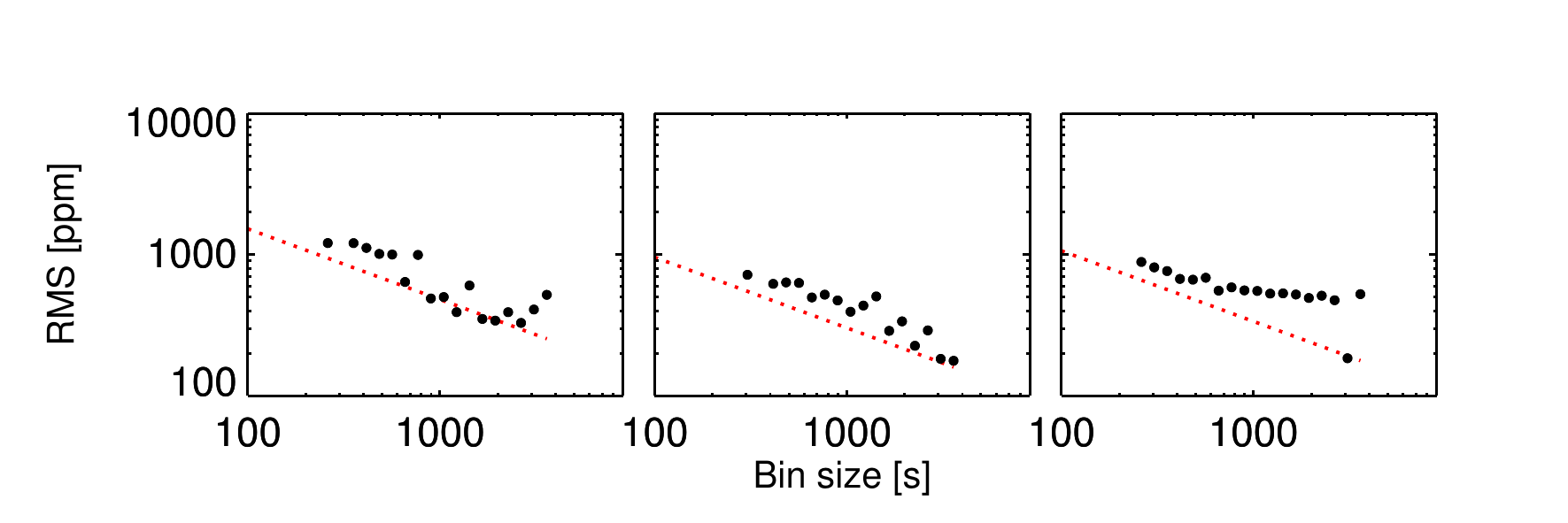}
\includegraphics[width=\columnwidth]{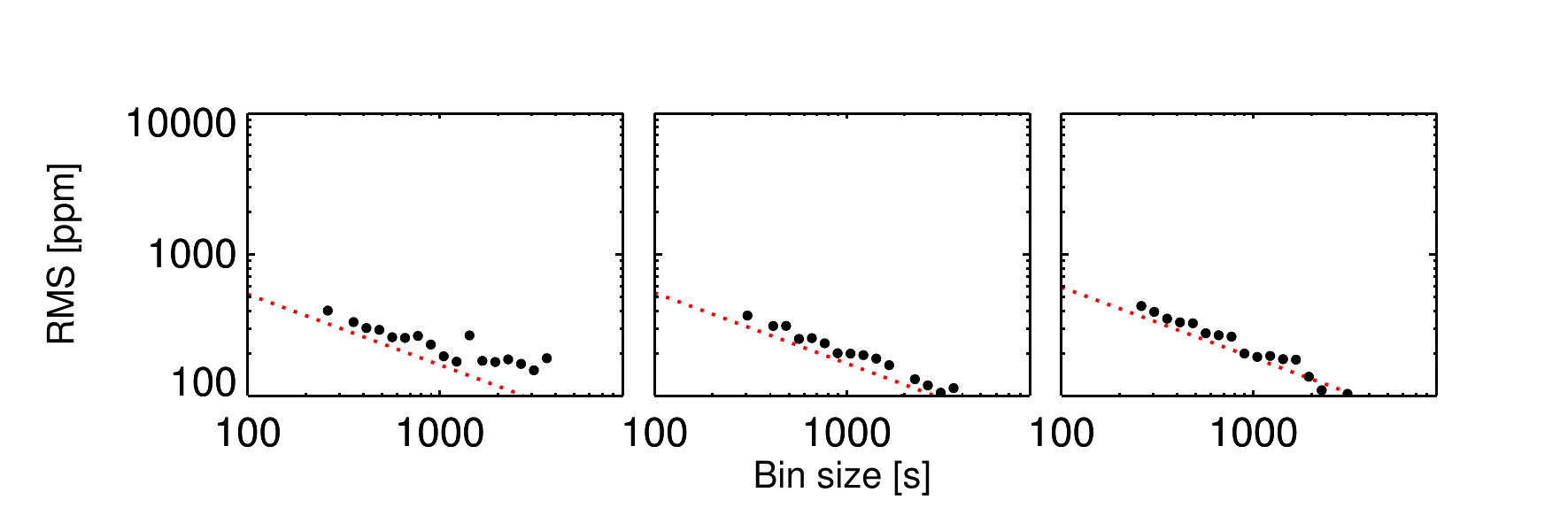}
\end{center}
\vspace{-0.1in}
\caption{\textit{Spitzer} 3.6 $\mu$m phase curves with BLISS mapping only (top-left panel) and additionally with PRF's FWHM (top-right panel).  The photometric residual RMS, in parts per million (ppm), is shown in the bottom panels as a function for each AOR.  In a purely Poisson-limited regime, we would expect to see the residual noise to decrease with the number of binned points ($N$) as $1/\sqrt{N}$.  Instead, we see non-monotonic behaviour with $N$.  With our updated baseline model, we find the residual correlated noise contribution to be nominal.}
\label{fig:tests}
\end{figure*}
\begin{figure*}
\begin{centering}
\includegraphics[width=0.65\columnwidth]{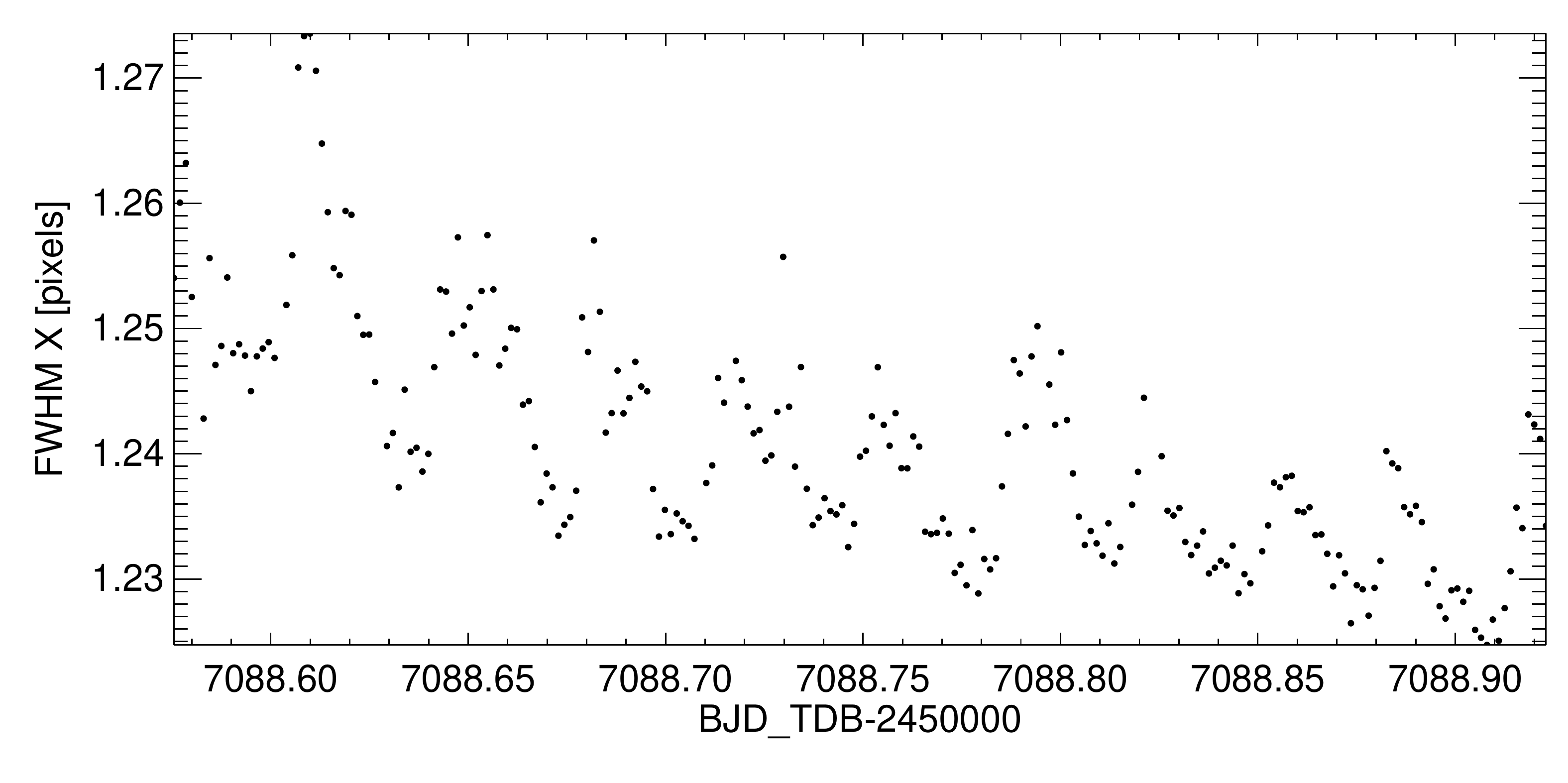}
\includegraphics[width=0.65\columnwidth]{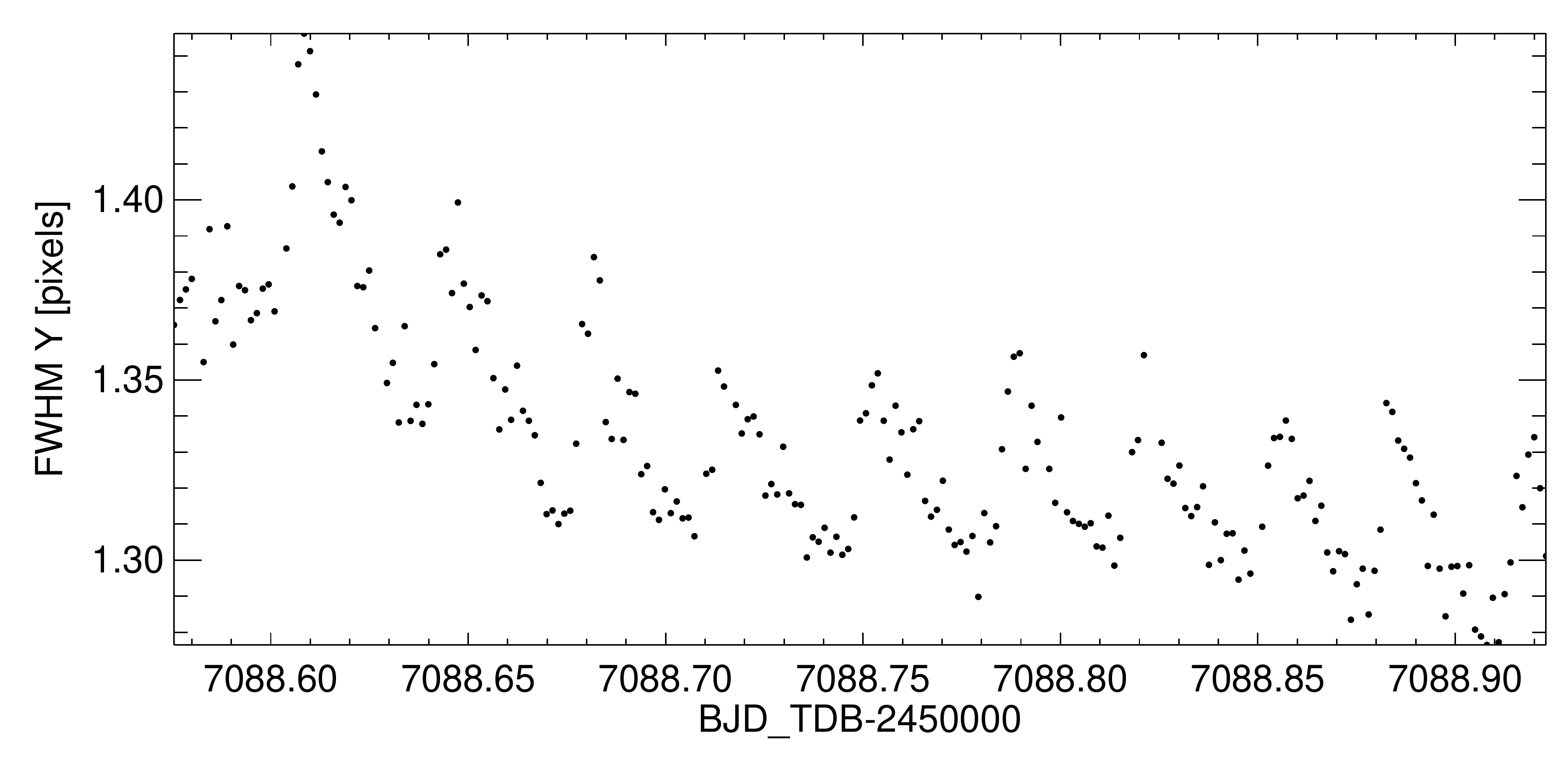}
\includegraphics[width=0.65\columnwidth]{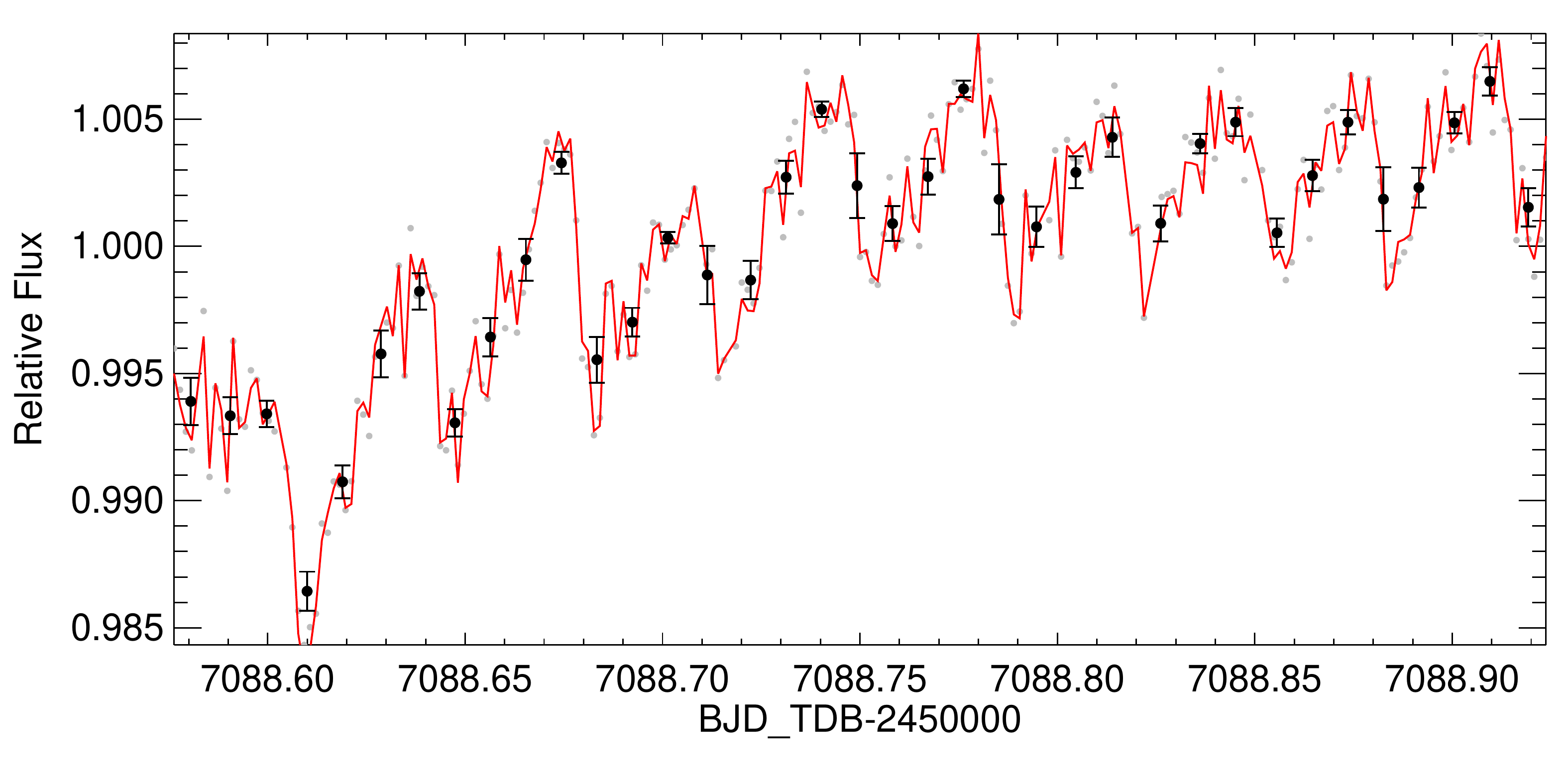}
\includegraphics[width=0.65\columnwidth]{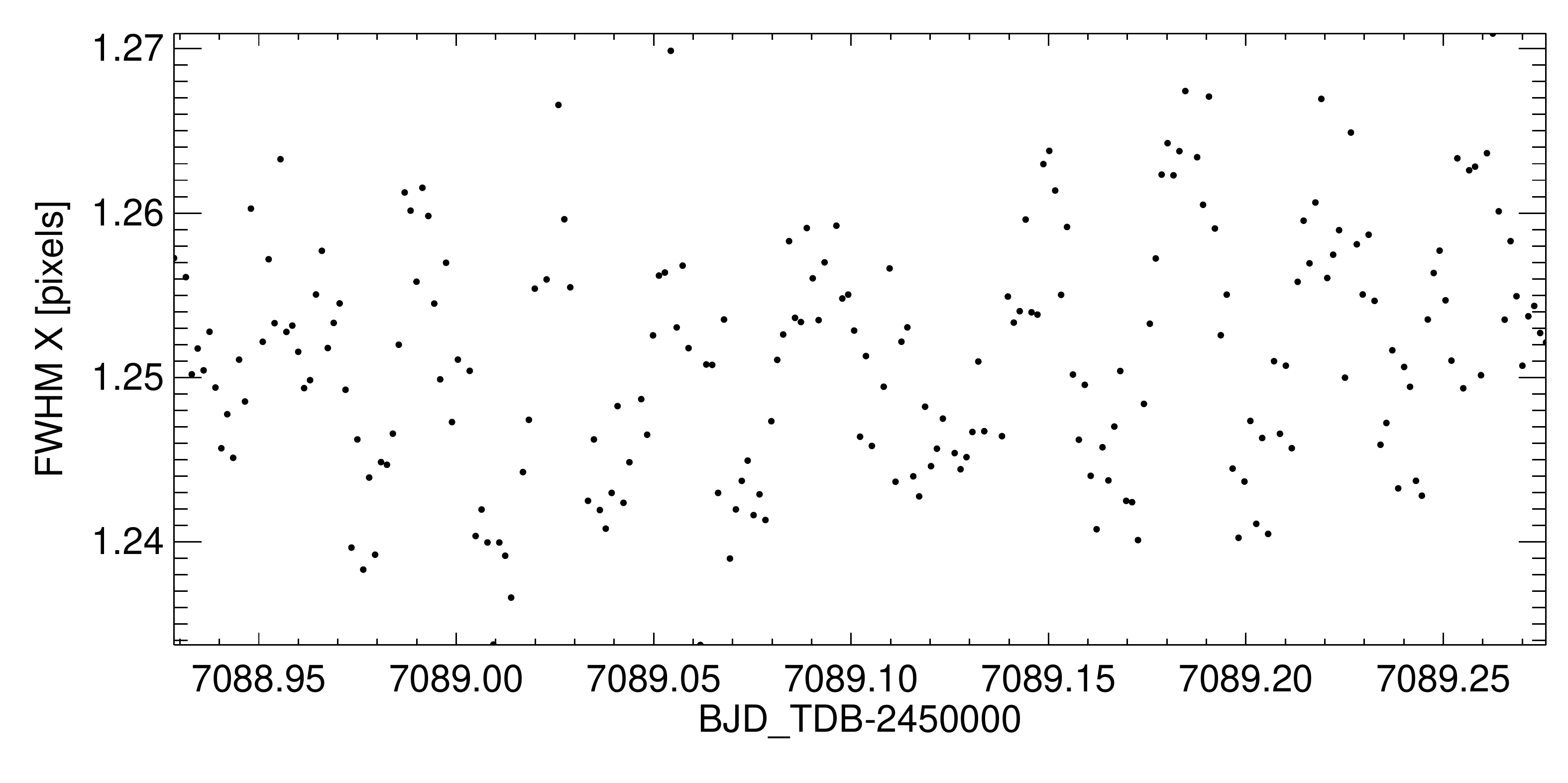}
\includegraphics[width=0.65\columnwidth]{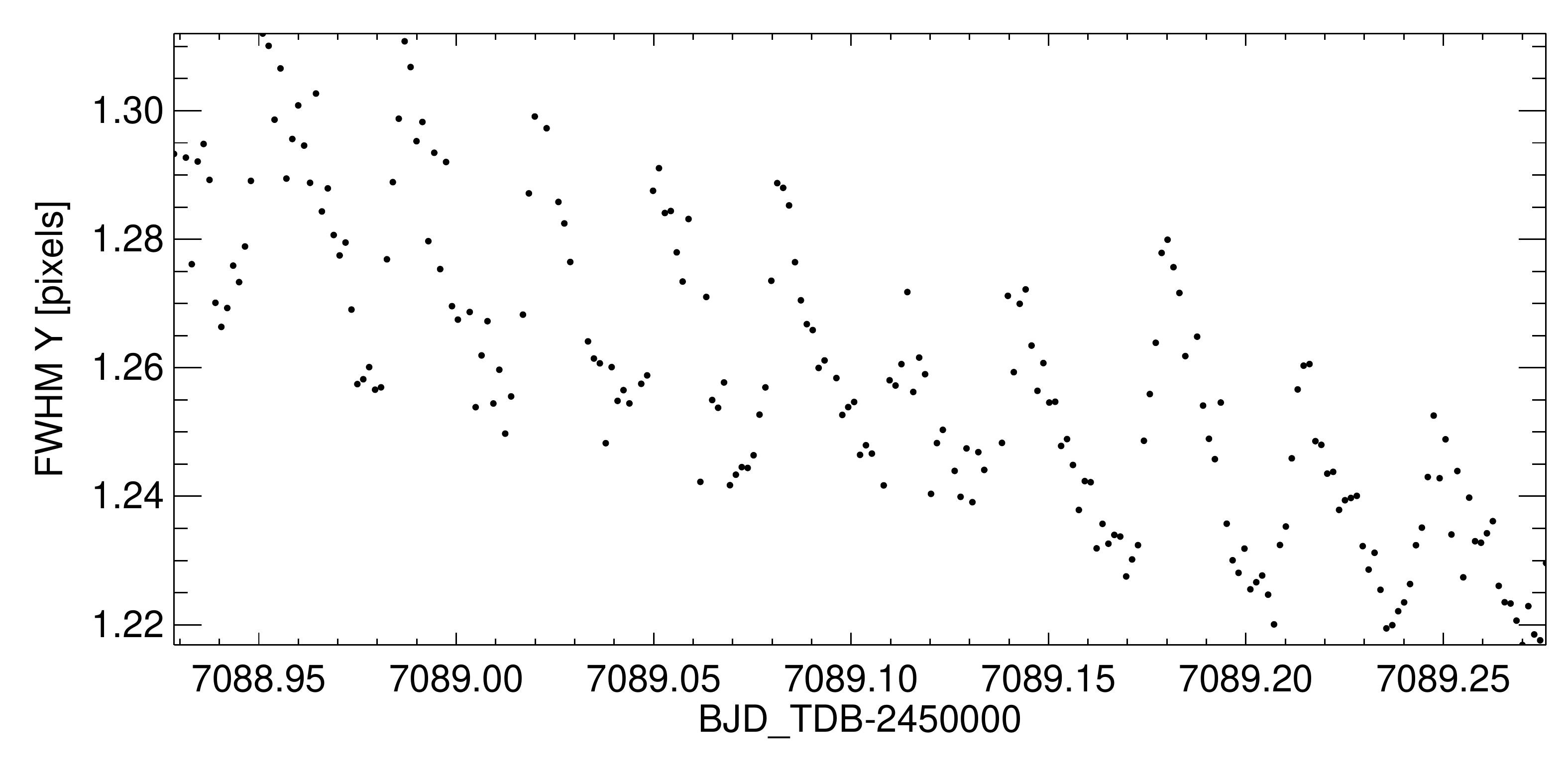}
\includegraphics[width=0.65\columnwidth]{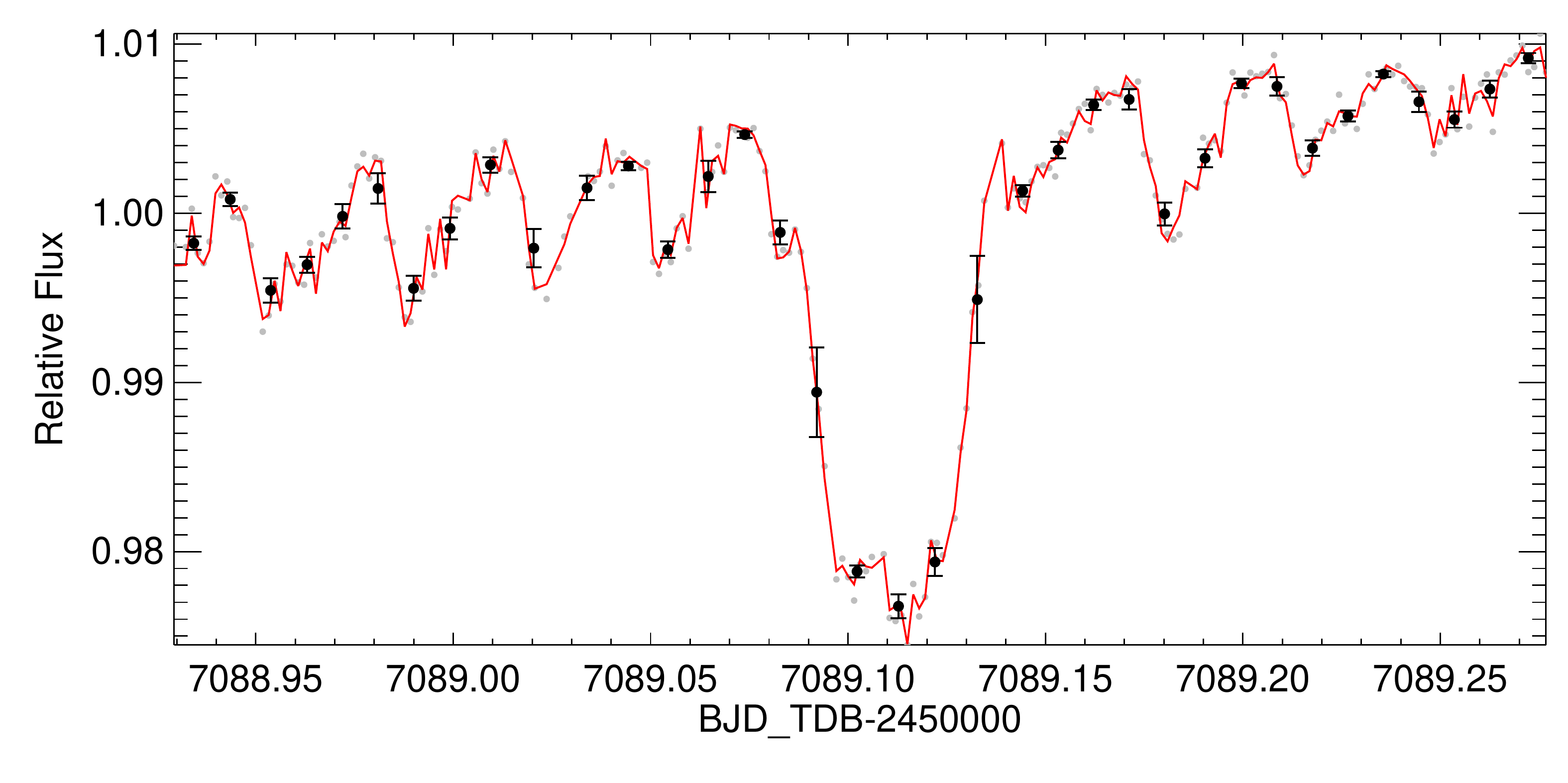}
\includegraphics[width=0.65\columnwidth]{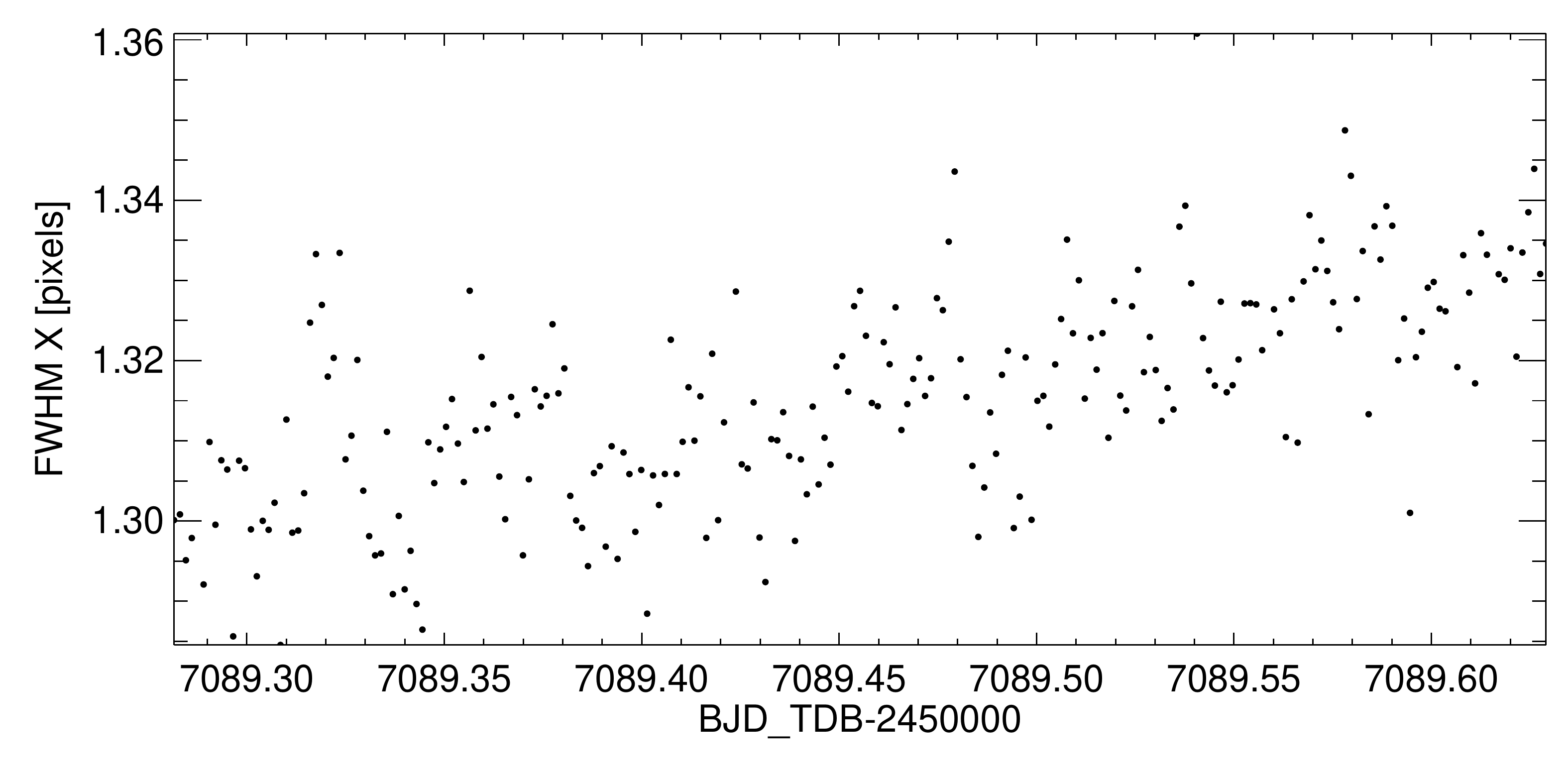}
\includegraphics[width=0.65\columnwidth]{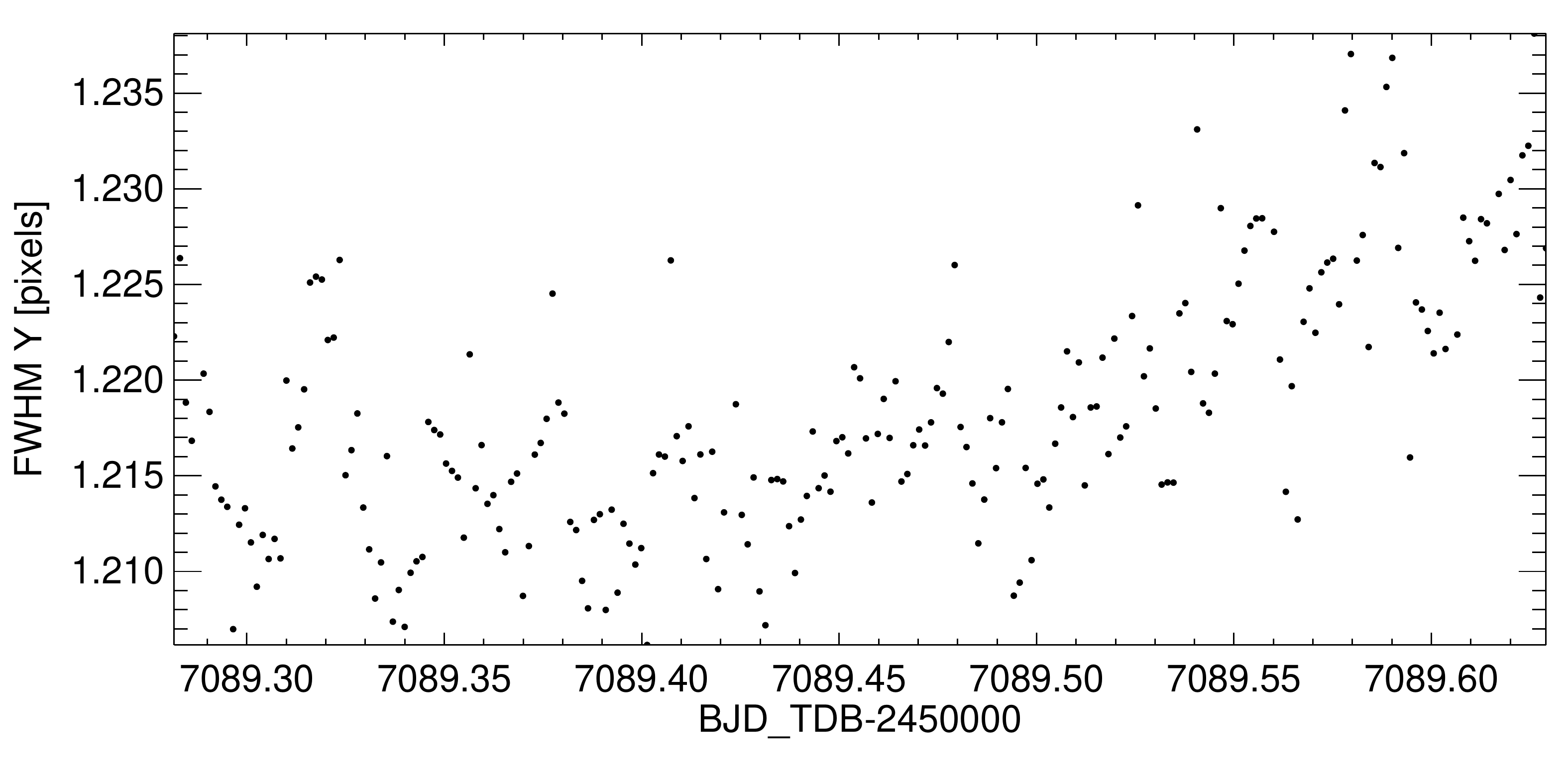}
\includegraphics[width=0.65\columnwidth]{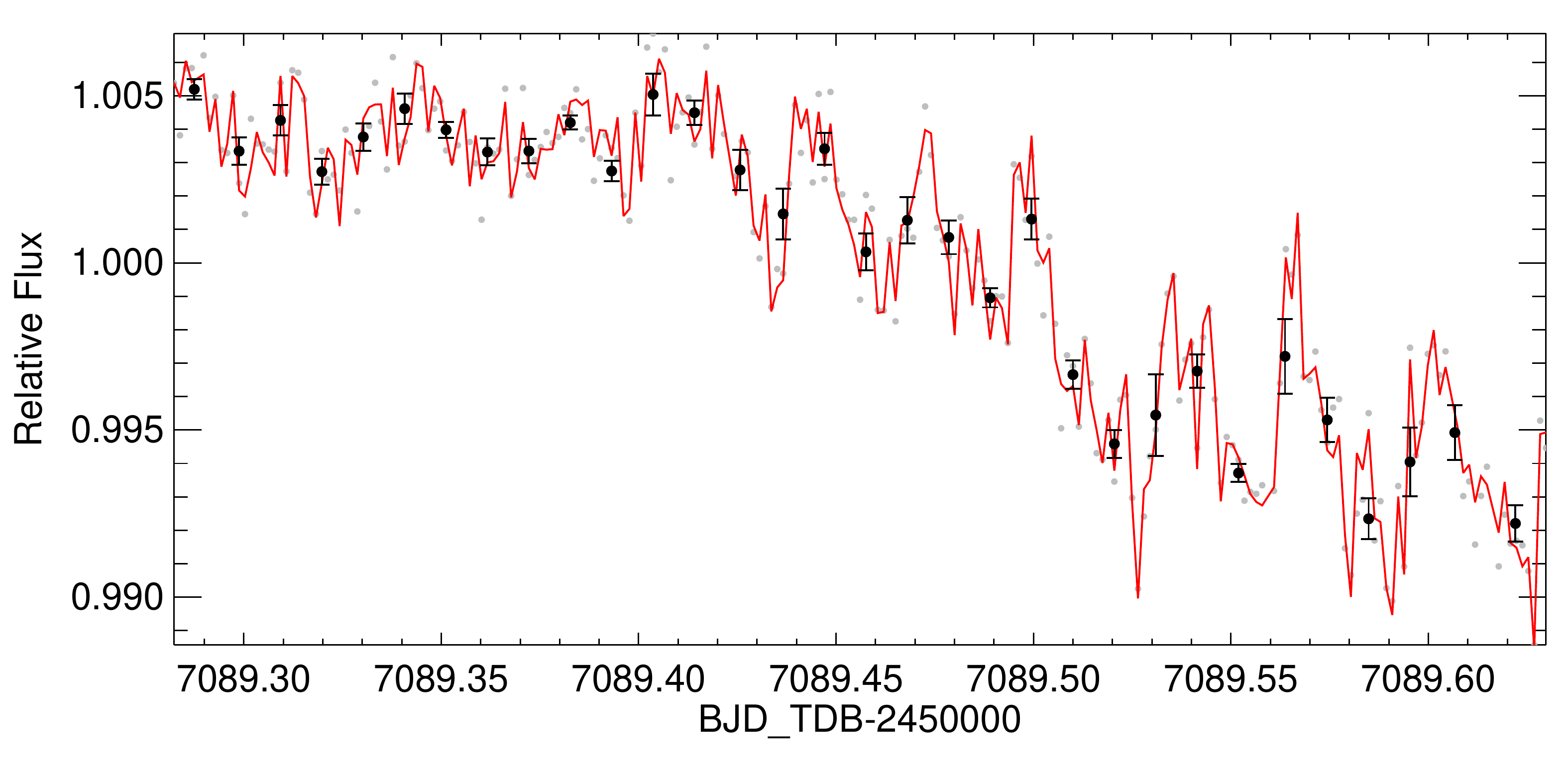}
\caption{Left and middle panels: PRF-FWHM; Right panels: raw flux data with the model superimposed for the three AORs of the first 3.6$\mu$m phase-curve.}
\label{fig:aors}
\end{centering}
\end{figure*}

\subsection{Theoretical motivation}

The near-infrared phase curves of \cite{2014Stevenson} were analyzed by \cite{2015Kataria}, who used the GCM described in \cite{2009Showman} to compute three-dimensional profiles of temperature and velocity, as well as multi-wavelength phase curves and emission spectra at different orbital phases.  The \cite{2009Showman} computational setup uses the \texttt{MITgcm} combined with two-stream radiative transfer and k-distribution opacities under the correlated-k approximation.  The computed dayside emission spectrum of \cite{2015Kataria} produces a reasonable fit to the measured dayside emission spectrum of \cite{2014Stevenson}, but over- or under-predicts the fluxes at other orbital phases.  Specifically, the model nightside emission spectrum is too bright compared to the measured one.  Clouds were mentioned as a possibility for reconciling models with data \citep{2015Kataria,2017Stevenson} but their impact on the atmospheric structure and observable spectra was not explored further. Together with the availability of additional data from \cite{2017Stevenson}, these facts motivate a revisiting of, and second opinion on, the GCM work associated with WASP-43b. 

It has been previously demonstrated that GCM outputs on hot Jupiters (velocity, temperature and hence fluxes) are uncertain by several tens of percent \citep{2011aHeng}.  This uncertainty arises from a tension between the need for computational feasibility and the desire for physical accuracy.  GCMs are very sensitive to choices of grid, computational methods (\citealt{2014Polichtchouk}) and how, for example, turbulence and eddy viscosity are represented at the subgrid scales (\citealt{2011aHeng}). These difficulties imply that it is good scientific practice for the same exoplanet to be simulated by more than one group using different GCMs with different algorithms, choice of grid, etc \citep{2015Heng}.  To this end, we propose to use our recently constructed GCM, \texttt{THOR}\footnote{\texttt{THOR} is an open-source software designed to run on Graphics Processing Units (GPUs): https://github.com/exoclime/THOR or https://bitbucket.org/jmmendonca/thor.} \citep{2016Mendoncab}, to simulate the atmosphere of WASP-43b and confront our simulated output with the data from both \cite{2014Stevenson} and \cite{2017Stevenson}.

\begin{figure*}

\begin{centering}
\includegraphics[width=1.0\columnwidth]{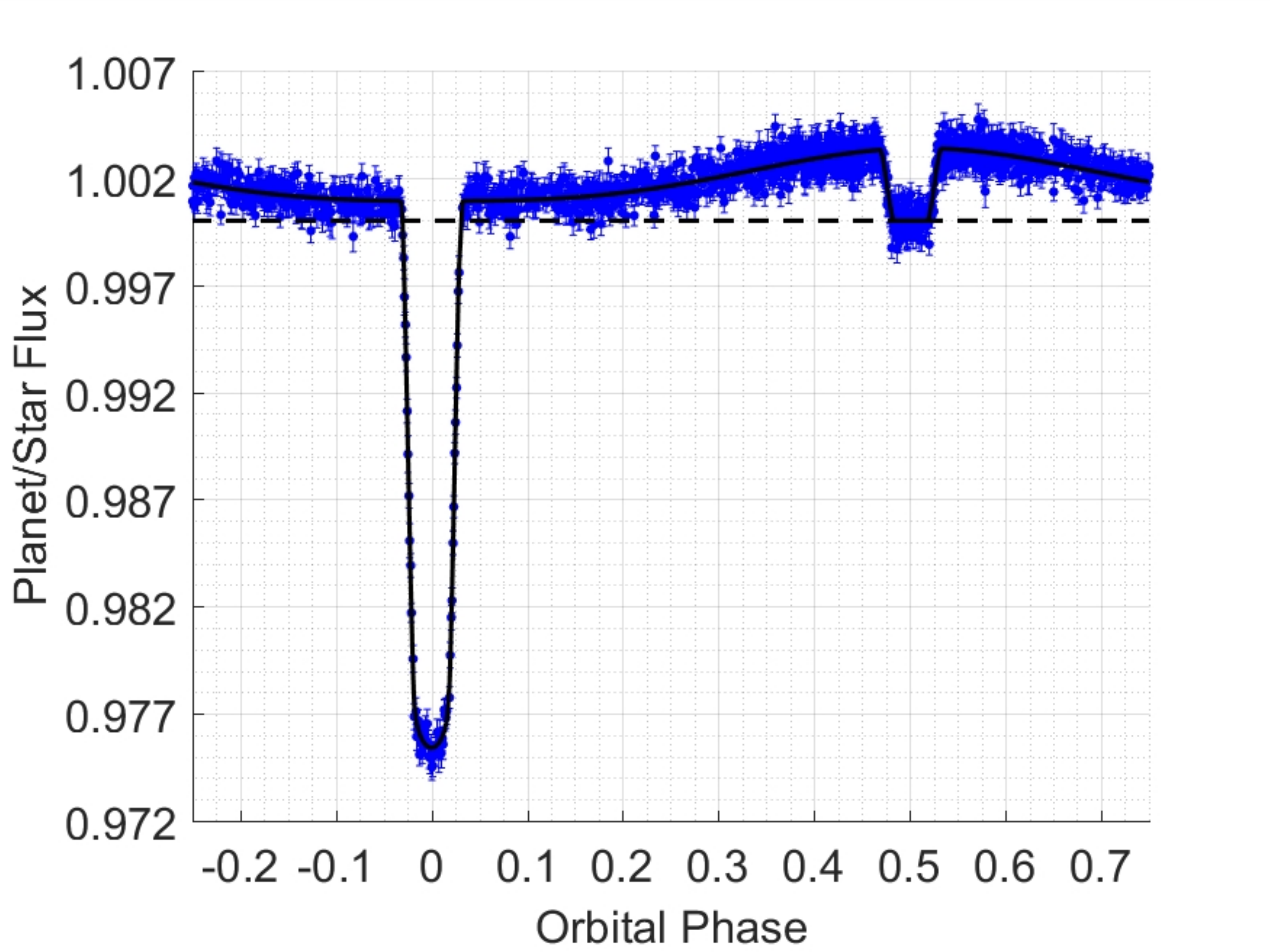}
\includegraphics[width=1.0\columnwidth]{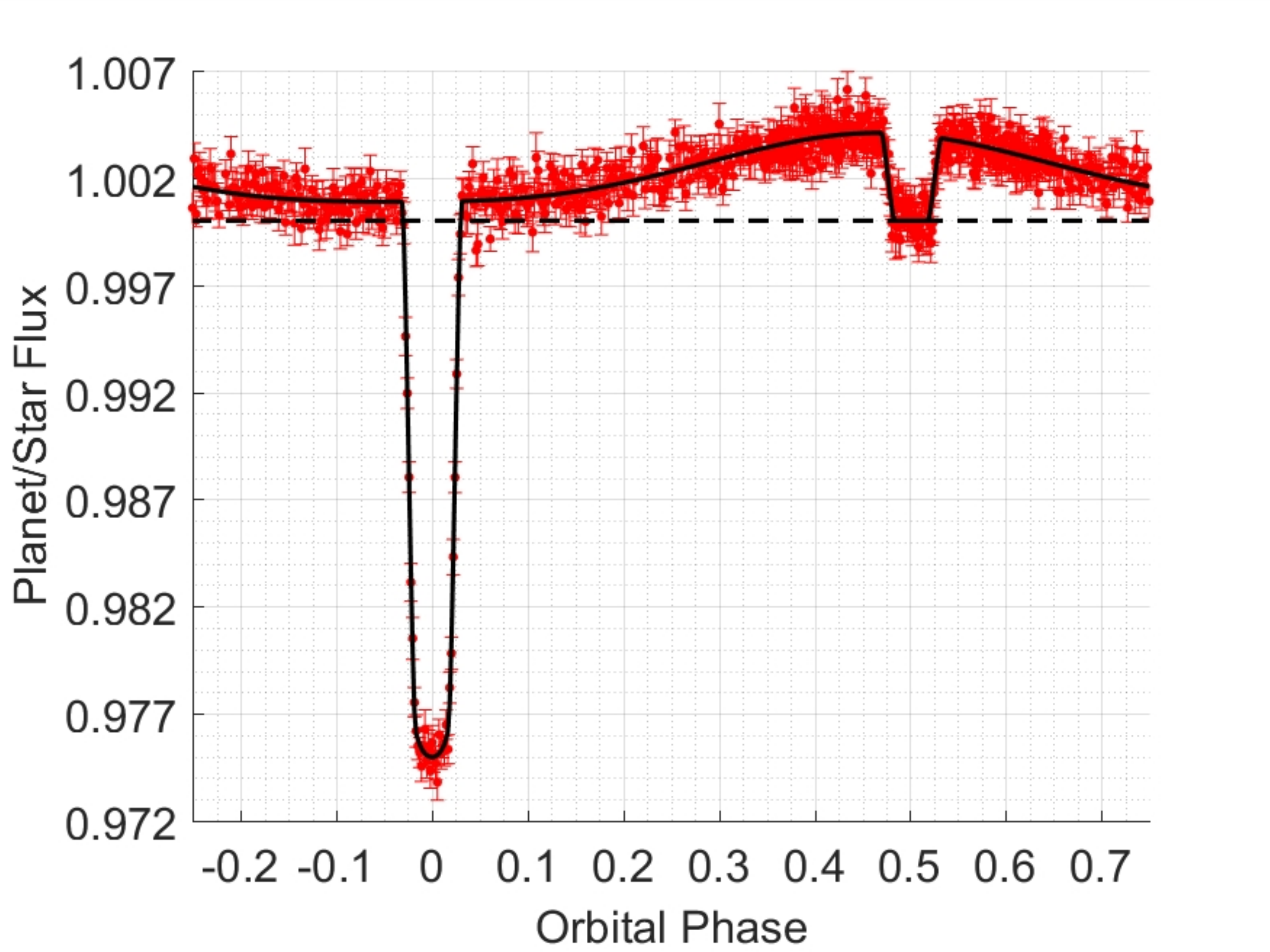}
\end{centering}
\caption{Combined spectroscopic phase curves of WASP-43b at 3.6 $\mu$m (left panel) and 4.5 $\mu$m (right panel). The solid line indicates the best-fit model to the observational data points. The values were normalized with respect to the stellar flux estimated during the secondary eclipse (orbital phase equal to 0.5)}
\label{fig:spitzer}
\end{figure*}

Besides providing the exoplanet community with a second opinion on the GCM of WASP-43b, we also include a simple parametrization of clouds in our \texttt{THOR} GCMs to explore the influence of clouds on the phase curves and emission spectra.  Like all other GCM studies, we will make specific choices for simplicity versus sophistication, which will contribute to a published hierarchy of GCMs on WASP-43b that is necessary for achieving understanding of physical processes.  The details of our computational setup are described in \S\ref{sec:model}.  

Thus, the other part of the present study is devoted to presenting new simulations of WASP-43b's atmosphere using a new GCM that has never been used to simulate this hot Jupiter.  While \cite{2015Kataria} focused mainly on comparing the outcomes of GCMs with $1\times$ and $5\times$ solar composition, we choose to focus on exploring if the properties of a cloud deck and the presence of enhanced carbon dioxide levels may be constrained by the data.  Thus, the two studies are complementary in their exploration of the physics.

\subsection{Structure of the present study}

\S\ref{sec:spitzer} is devoted to providing the details of our re-analysis of the \cite{2017Stevenson} \textit{Spitzer} data, and our justification for why the \textit{HST} data of \cite{2014Stevenson} does not require re-analysis.  \S\ref{sec:model} contains a detailed description of our \texttt{THOR} GCM and a comparison of it to other GCMs.  Our results are presented in \S\ref{sec:results}, including multi-faceted comparisons of models with data.  Our conclusions, as well as prospects for future work, are described in \S\ref{sec:conclu}.

\section{\textit{Spitzer} data reduction and analysis}
\label{sec:spitzer}

\subsection{Reanalysis of Spitzer phase-curve data}
In this section, we justify our re-analysis of the \cite{2017Stevenson} 3.6 $\mu$m phase curves by using a modified version of the BLISS mapping (BM) technique that was used by these authors to mitigate the correlated noise associated with intra-pixel sensitivity. In our photometric baseline model, we complement the BM algorithm with a linear function of the Point Response Function (PRF) Full Width at Half-Maximum (FWHM). We find this addition to yield a dramatic improvement on the photometric residuals (Figure \ref{fig:tests}, bottom panels). Next, we explain in greater detail the procedure we follow.

We downloaded WASP-43b archival IRAC data from the Spitzer Heritage Archive (\url{http://sha.ipac.caltech.edu}).  The data consists of one WASP-43b phase-curve at 4.5 $\mu$m (split in 3 AOR, PID 10169) and two at 3.6 $\mu$m (split in 3 and 2 AOR respectively, PID 11001) \citep{2017Stevenson}. The reduction and analysis of these datasets are similar to \citet{2016Demoryb}. We model the IRAC intra-pixel sensitivity \citep{2016Ingalls} using a modified implementation of the BLISS (BiLinearly-Interpolated Sub-pixel Sensitivity) mapping algorithm (\citealt{2012Stevenson}).

In addition to the BLISS mapping used in \citet{2017Stevenson}, our baseline model includes the PRF's FWHM along the $x$ and $y$ axes, which significantly reduces the level of correlated noise as shown in previous studies (\citealt{2014Lanotte}; \citealt{2016Demorya}; \citealt{2016Demoryb}; \citealt{2017Gillon}). In the following, we compare the two baseline models: one with bliss-mapping (BM) alone and one that combines BM and the PRF FWHM (BM+FWHM). The reason for the improvement when using the BM+FWHM model is that the point response function shape evolves with time and its properties are not accounted for by the BLISS algorithm alone. The baseline model does not include time-dependent parameters, such as a ramp. It can be seen from Fig. \ref{fig:aors} that the raw data does not exhibit a ramp-like feature. Our implementation of this baseline model is included in a Markov Chain Monte Carlo (MCMC) framework already presented in the literature (\citealt{2012Gillon}). We run two chains of 200,000 steps each to determine the phase-curve properties at 3.6 and 4.5 $\mu$m based on the entire dataset described in the paragraph above (Fig. \ref{fig:spitzer}).

We compare in Figure \ref{fig:tests} our analysis of the first 3.6 $\mu$m phase-curve to the one of \citet{2017Stevenson}. Contrary to this study, we do not detect systematic features connected to the uneven sampling of the target on the detector. An examination of the light-curve residuals reveals nominal contribution from correlated noise. We use a simple baseline model comparison for this phase-curve, using the Bayesian Information Criterion (BIC, see \citealt{1978Schwarz}). We find BIC values of 2626 and 859 with BM alone and BM+FWHM respectively, which favours the addition of the FWHM parameters in the baseline model for this dataset. In comparing both 3.6 $\mu$m phase-curves, we find  amplitudes of 2395$\pm$190 and 2719$\pm$200ppm for the two 3.6 $\mu$m phase curves, compared to the previously published 2440$\pm$230 and 3380$\pm$110 ppm values. We do not detect the $\sim$300 ppm systematic feature previously reported in the second 3.6 $\mu$m phase-curve. 

From our BM+FWHM baseline model, we compute a photon-limited precision of 535 ppm per 123s exposure time at 3.6 $\mu$m. The BM-only baseline model yields a precision of 955 ppm per 123s exposure time. The corresponding photon-limited precision is 455 ppm per exposure for these data. Our results are typical of IRAC 3.6 $\mu$m photometric performance ($\sim$15$\%$ above the photon noise limit).

This test demonstrates that the impact of slight non-repeatability in the positioning of the star on the detector can be significantly mitigated with an appropriate baseline model.

We also perform an analysis of the Spitzer 4.5 $\mu$m data with the same photometric baseline model for which we find a phase-curve amplitude of 3258$\pm$250 ppm.

\subsection{HST-WFC3 data}
\textit{Wide Field Camera 3 (WFC3)} is known to perform well down to the photon noise limit in spatial scan mode \citep{2013Deming}. We therefore elect to use the published data as is.

\section{The \texttt{THOR} Global Circulation Model}
\label{sec:model}

\begin{table}
\begin{center}
\caption{Input parameters used in the reference GCM simulations of WASP-43b.}
\begin{tabular}{ | l | l | l |}
\hline
 Parameters & Value Adopted & Units  \\ \hline \hline
 Star Temperature & 4520 & K \\ \hline
 Planet distance & 0.015 & AU \\ \hline
 Mean Radius & 72427 & km \\ \hline
 Gravity & 47.0 & m/s$^2$ \\ \hline
 Gas constant & 3714 & J/K/kg \\ \hline
 Specific heat & 13000 & J/K/kg \\ \hline
 Bond albedo & 0.18 & - \\ \hline
 Highest pressure & $\approx$100 & bar \\ \hline
 Interior flux & $\approx$ 50 & kW/s$^2$ \\ \hline
 Rotation rate & 9.09$\times10^{-5}$ & s$^{-1}$ \\ \hline
 Orbit inclination & 0 & deg \\ \hline
 Orbit eccentricity & 0 & deg \\ \hline
 \end{tabular}
\end{center}
\label{tab:model}
\end{table} 

\subsection{Computational setup in the context of previous GCMs}

Our \texttt{THOR} GCM was developed, from scratch, to solve the non-hydrostatic Euler equations on an icosahedral grid \citep{2016Mendoncab}.  \texttt{THOR} has been demonstrated to reproduce the standard benchmark tests for Earth and exoplanet GCMs suggested by \cite{2011aHeng}. By comparison, most of the GCMs published in the exoplanet literature solve a reduced set of equations known as the primitive equations of meteorology, which assume hydrostatic equilibrium, a thin atmosphere and neglect radial Coriolis terms (e.g., \citealt{2014Mayne} and \citealt{2015Heng} for review papers on common dynamical approximations). Read \cite{2016Mendoncab} to learn more about the numerical and physical robustness of THOR.  The GCM of \cite{2015Kataria} performs multi-wavelength radiative transfer, but solves the primitive equations.  In our GCM, we use a different approach: included a simpler ``double-grey" radiative transfer (see appendix \ref{apxd:rad_tr}), where radiation is split into the optical/visible (from the star) and infrared (from the exoplanet) wavebands, but solved the non-hydrostatic Euler equations. For each waveband, one needs to specify a mean opacity.  The optical/visible and infrared opacities are set to be 0.025 cm$^2$g$^{-1}$ and 0.05 cm$^2$g$^{-1}$, respectively, which correspond to a photon deposition depth (in the optical/visible) and a photospheric pressure (in the infrared) $\sim 100$ mbar consistent with the one obtained in \cite{2014Stevenson}. Our simplified scheme is very efficient and similar to the ``double-grey" radiative transfer used in \cite{2011Heng}, who showed that the resulting global structure of the hot Jovian atmosphere is qualitatively similar to that obtained by \cite{2009Showman} using multi-wavelength radiative transfer. Our simplified scheme also captures quantitatively the longitudinal temperature distribution from \cite{2015Kataria} (compare Fig. 7 in \cite{2015Kataria} with our Fig. \ref{fig:ref_results_u_temp}), which is important to interpret the observational data along the longitude. However, using our simple radiative transfer we do not represent scattering or the multiple wavelength optical structure of the atmosphere, which reduces the accuracy of the heating/cooling rates profiles in the atmosphere. This limitation has an important impact mainly in the deep atmosphere ($>$1 bar) as it is discussed later on the phase-curves around 1 $\mu$m that are probing the deep levels (read section \ref{sec:const-curve}). Changes in chemical abundances across the atmosphere is also neglected in this approach. We are working currently on the development of a flexible multi-wavelength radiative transfer scheme for THOR with the same level of sophistication as in \cite{2009Showman} and \cite{2014Amundsen}, which will help us refine the results found in this work.

Following \cite{2015Kataria}, we assume WASP-43b to be tidally locked.  The input parameters for our GCMs are listed in Table \ref{tab:model}, and largely follow what were assumed by \cite{2015Kataria} in order to facilitate comparison.  Each GCM run was started from an isothermal (1400 K) state of rest and integrated for 7500 Earth days, with a timestep of 300 seconds, until a statistical steady state of the deep atmosphere thermal structure was obtained. This long integration is important to avoid the results being biased towards the set initial conditions.  The horizontal resolution used is about 4 degrees on a sphere.  The subgrid scale dissipation is represented by a fourth-order hyperdiffusion and a 3D divergence damping (\citealt{2016Mendoncab}) with the same diffusion time-scale of 940 seconds; see \cite{2011aHeng} for a discussion of hyperdiffusion on tidally locked exoplanets. Our model atmospheres consist of 40 discrete layers with pressures ranging from about 100 bar to 0.01 mbar.  A convective adjustment scheme is used, which consists of mixing vertically the entropy in the atmosphere when the lapse rate becomes super-adiabatic, while conserving the total enthalpy of the unstable atmospheric column \citep{1965Manabe,2016Mendoncab}. We implement a correction of the cosine of the zenith angle to represent the effective path length (see \citealt{2006Li} and \citealt{2016Mendoncaa}) that is affected by the spherical geometry of the planet.  We include a bond albedo of 0.18 that was estimated in \citealt{2014Stevenson}. The flux coming from the planet's interior was calculated ($\approx$ 50 kW$/$m$^2$) to represent an equilibrium temperature consistent with the observational data (\citealt{2014Stevenson} and our new \textit{Spitzer} re-analysis).  

\subsection{Treatment of clouds in GCMs}

Cloud distribution and composition in hot Jupiters continue to be an active topic of exploration.  Using the \texttt{MITgcm}, \cite{2016Parmentier} modeled purely absorbing aerosols with tracers that included a treatment of their size-dependent terminal velocity.  Using the \texttt{FMS} GCM, \cite{2016Oreshenko} generalized two-stream radiative transfer to include scattering and overlaid condensation curves on the simulated three-dimensional temperature profile to approximate the spatial distribution of aerosols. In \cite{2016Lee} the first steps towards self-consistent simulations of clouds, radiation and 3D atmospheric circulation on hot Jupiters are described. This type of simulations will be important to help us improve our understanding on the atmospheric processes associated with cloud formation and transport in hot Jupiter planets.

In the current study, we make the simplest assumption and include clouds in our WASP-43b GCMs in the form of a constant, additional opacity covering the nightside of the planet. The physical assumptions behind such an approximation are that the cloud particles are large compared to the wavelengths of thermal emission and that the timescale for them to condense out of the atmospheric gas is short compared to any dynamical or radiative timescales. Using this approach, we also avoid choosing a cloud composition and particle size, which continue to be poorly known (e.g., \citealt{2016Parmentier}). A reflected-light phase-curve in the visible may help us in the future to constrain the composition, particle distribution, and spatial distribution of the clouds (see \cite{2013Marley} for a review on exoplanet clouds). A similar treatment of clouds is used in \cite{2013Dobbs-Dixon}, however, in our case, we set constrains on the cloud spatial distribution. The vertical cloud distribution is very sensitive to local temperature and vertical mixing (\citealt{2015Lee}). We simplify the unconstrained vertical distribution by assuming that the cloud extends from 100 mbar to 1 bar, and for its opacity to decay linearly with pressure above 100 mbar: $k_{cloud} = p\times5\times 10^{-4}$ cm$^2$g$^{-1}$, where $p$ is the pressure in mbar. The decrease of the absorption with altitude is a crude representation of the decrease of cloud density with altitude due to the settlement of the large cloud particles toward higher pressures. The model produces very similar results to those adopting a well mixed cloud structure, e.g., \cite{2016Parmentier}. Above 1 mbar, the cloud opacity is set to zero. A very high cloud top would be inconsistent with recent transmission spectroscopy results (e.g., \citealt{2014Kreidberg}), and, would replace the still detected water absorption feature from the nightside emission with a continuum emission. Our cloud structure blocks the radiation coming from the deepest layers and raises the photosphere in the nightside to roughly 10 mbar (defined here as the cloud top level).  We would like to highlight the possibility of other cloud solutions to represent the cloud cover in WASP-43b, but since the observational data is still very crude to constrain the cloud properties we keep this setting as simple as possible. The cloud cover is located on the nightside of the planet, with no interaction with the stellar-light. The clouds are positioned where $\cos\theta{_z}$ is negative and its density weighted by $|\cos\theta_z|^{0.2}$, where $\theta_z$ is the zenith angle, to avoid undesired sharp transitions in the thermal structure during the GCM simulations. Note that the use of this weighting function does not produce noticeable differences in the post-processing results explored later. We further explore the effects of shifting the cloud westward by 20 degrees, where the weighting function is then defined as $|\cos(\lambda+20^o)\cos\phi|^{0.2}$ with $\lambda$ being the longitude and $\phi$ the latitude. In this case, the stellar-light interacts with the cloud cover for zenith angles higher than $80^o$ but it does not significantly affect the atmospheric thermal structure as it is pointed out in the analysis of the phase-curves later in section \ref{subsec:mespectra}.

\subsection{Computing observables}

Upon obtaining the three-dimensional structure of temperature and pressure, we post-process this output to obtain multi-phase synthetic spectra and multi-wavelength phase curves.  The spectra are generated combining the radiative emission solution from \cite{2014Heng} and \cite{2017Malik} with the multiple-scattering solution to treat the stellar radiation from \cite{2015Mendonca}.  For these results we include the main absorbers in the infrared from two databases: HITEMP (\citealt{2010Rothman}) for H$_2$O, CO$_2$ and CO; HITRAN (\citealt{2013Rothman}) for CH$_4$, NH$_3$, HCN, C$_2$H$_2$ and the collision-induced absorption from H$_2$-H$_2$ and H$_2$-He (\citealt{2012Richard}). We have also included Rayleigh scattering by hydrogen molecules. The projected outgoing intensity at the top of the atmosphere is calculated for each geographical location of the observed hemisphere that moves with the orbital phase. The spectral resolution used was 3000 spectral bins covering a spectral range from $0.3$ $\mu$m to 10 cm.  The abundances in the atmosphere of WASP-43b are assumed to be in chemical equilibrium and were computed using \texttt{FastChem} (\citealt{2017Stock}). The stellar flux was interpolated from the PHOENIX model database (\citealt{1995Allard}; \citealt{2013Husser}). The post-processing tools take also into account the same 3D cloud structures with gray opacity as used in the GCM simulations to maintain consistency in the interpretation of the results.

\begin{figure*}
\begin{centering}
\subfigure[Zonal wind - no clouds]{
\includegraphics[width=0.65\columnwidth]{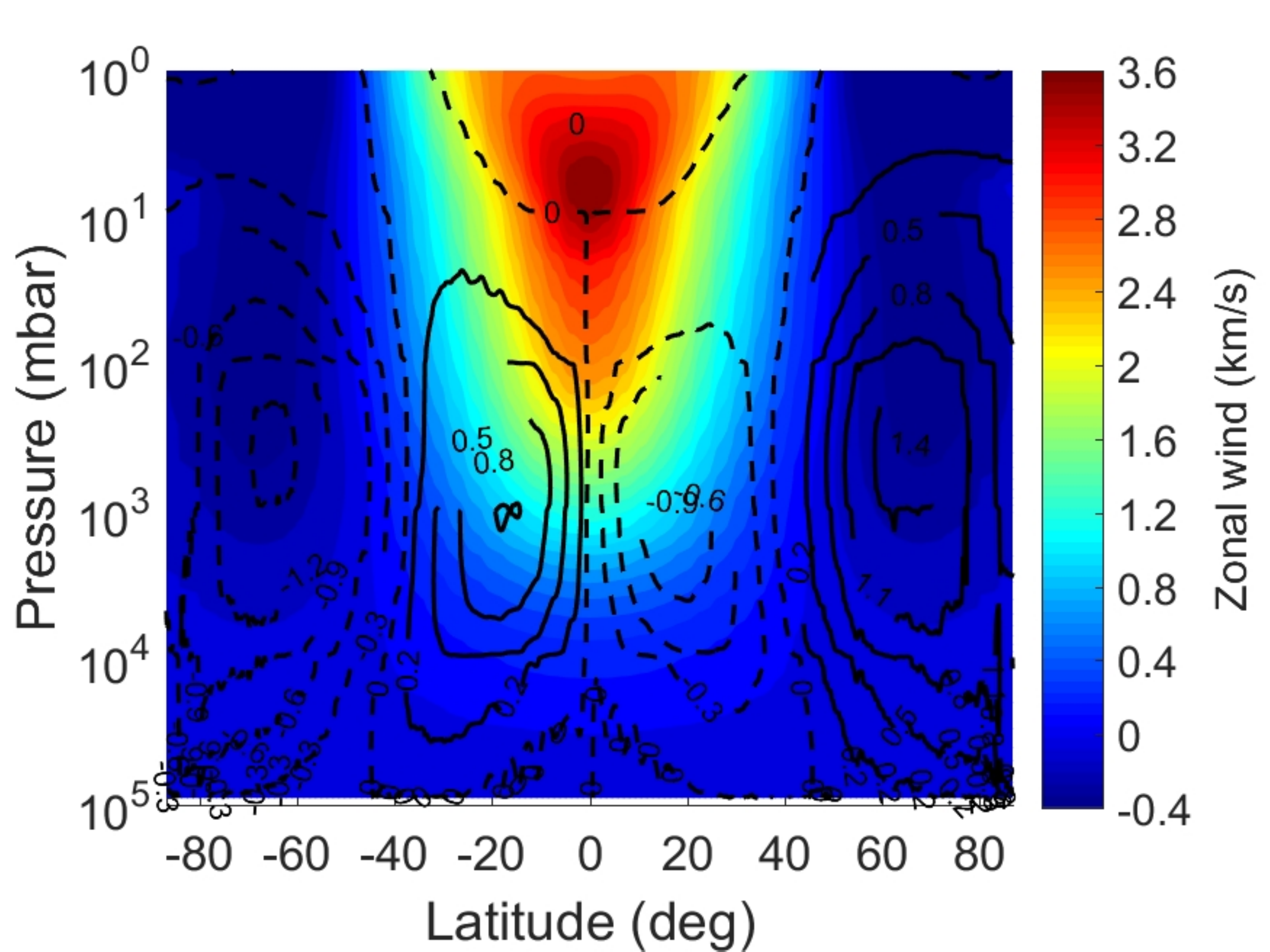}}
\subfigure[Temperature - no clouds]{
\includegraphics[width=0.65\columnwidth]{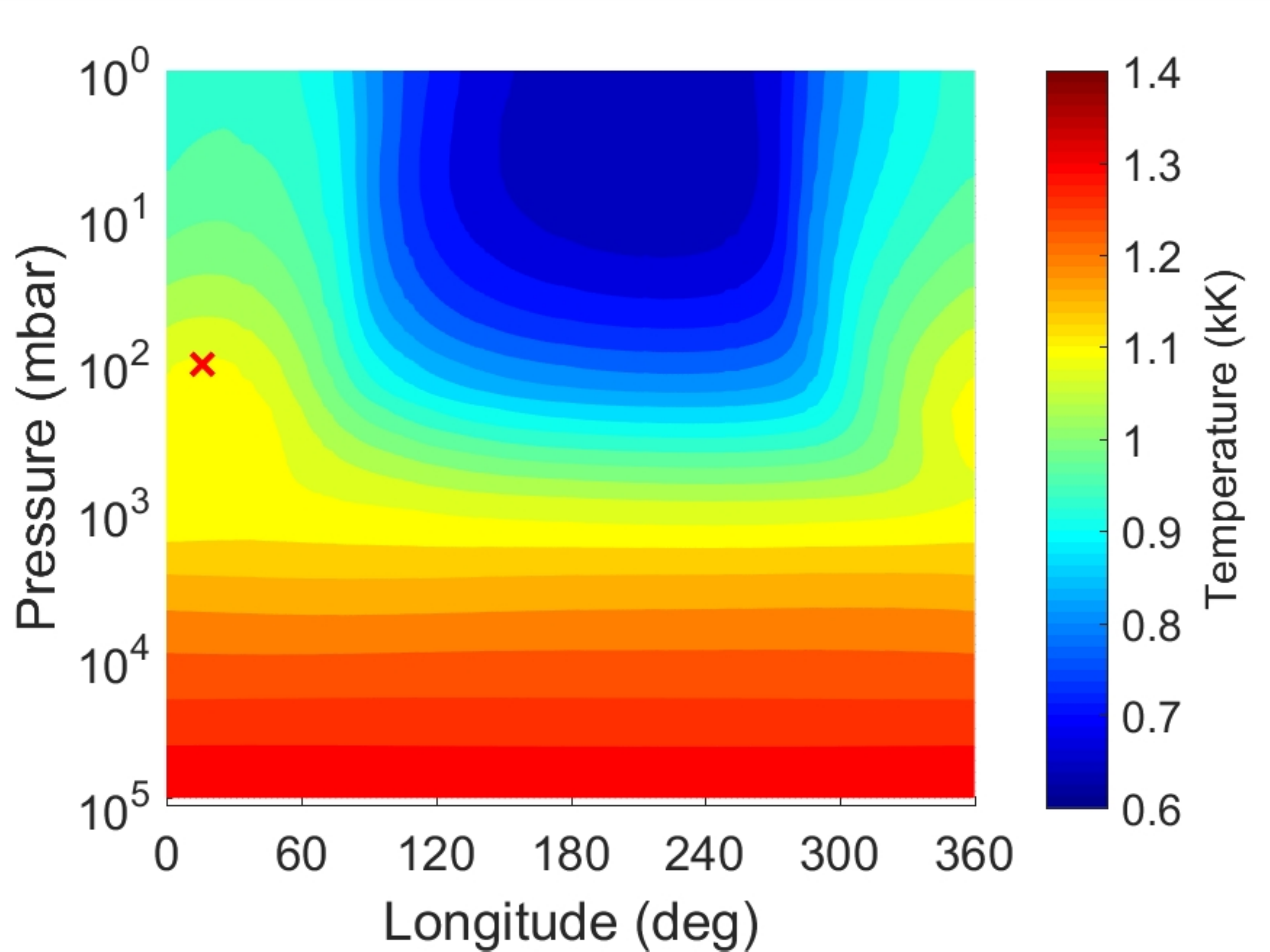}}
\subfigure[Temperature at 10 mbar - no clouds]{
\includegraphics[width=0.65\columnwidth]{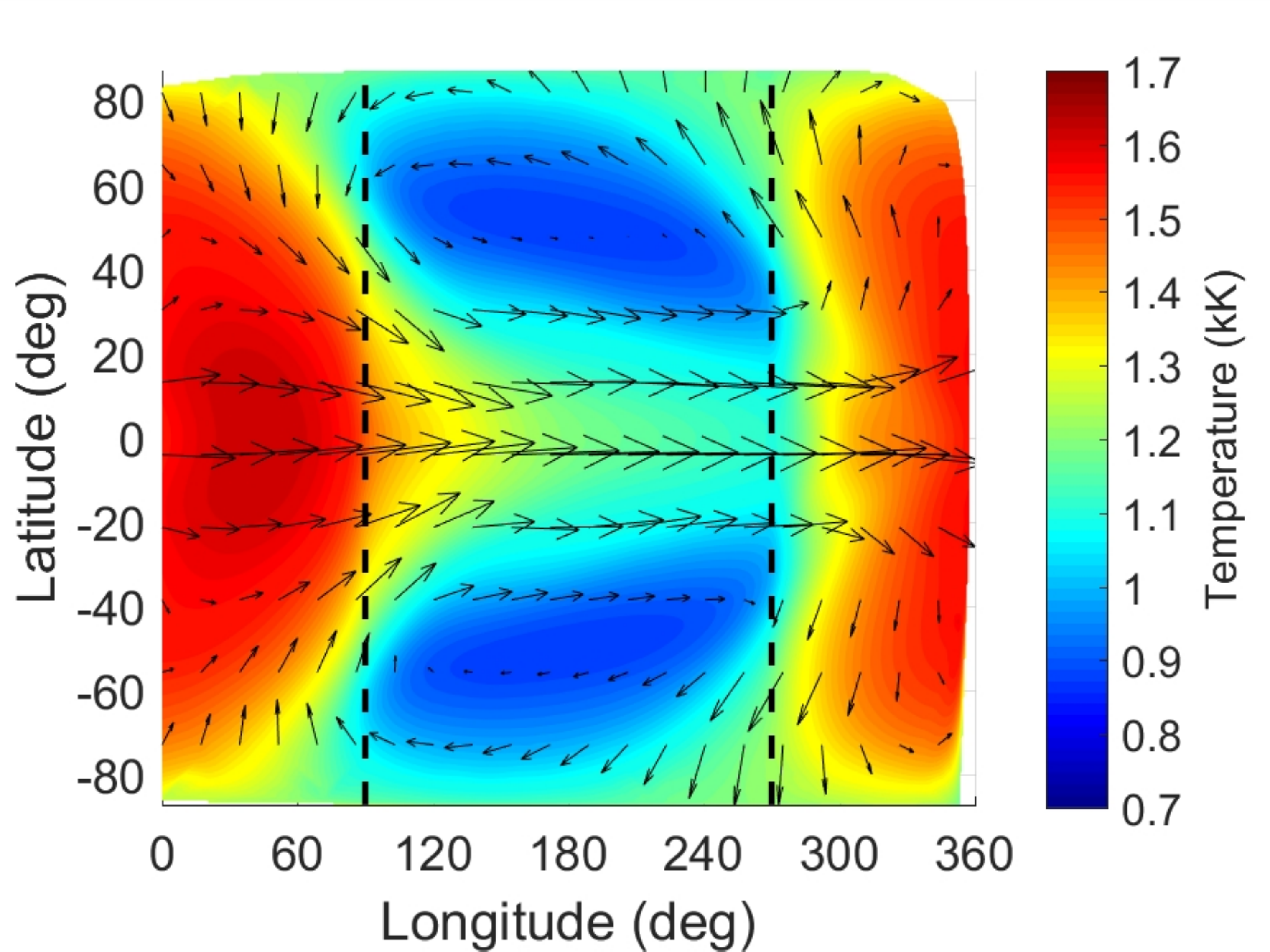}}
\subfigure[Zonal wind - with clouds]{
\includegraphics[width=0.65\columnwidth]{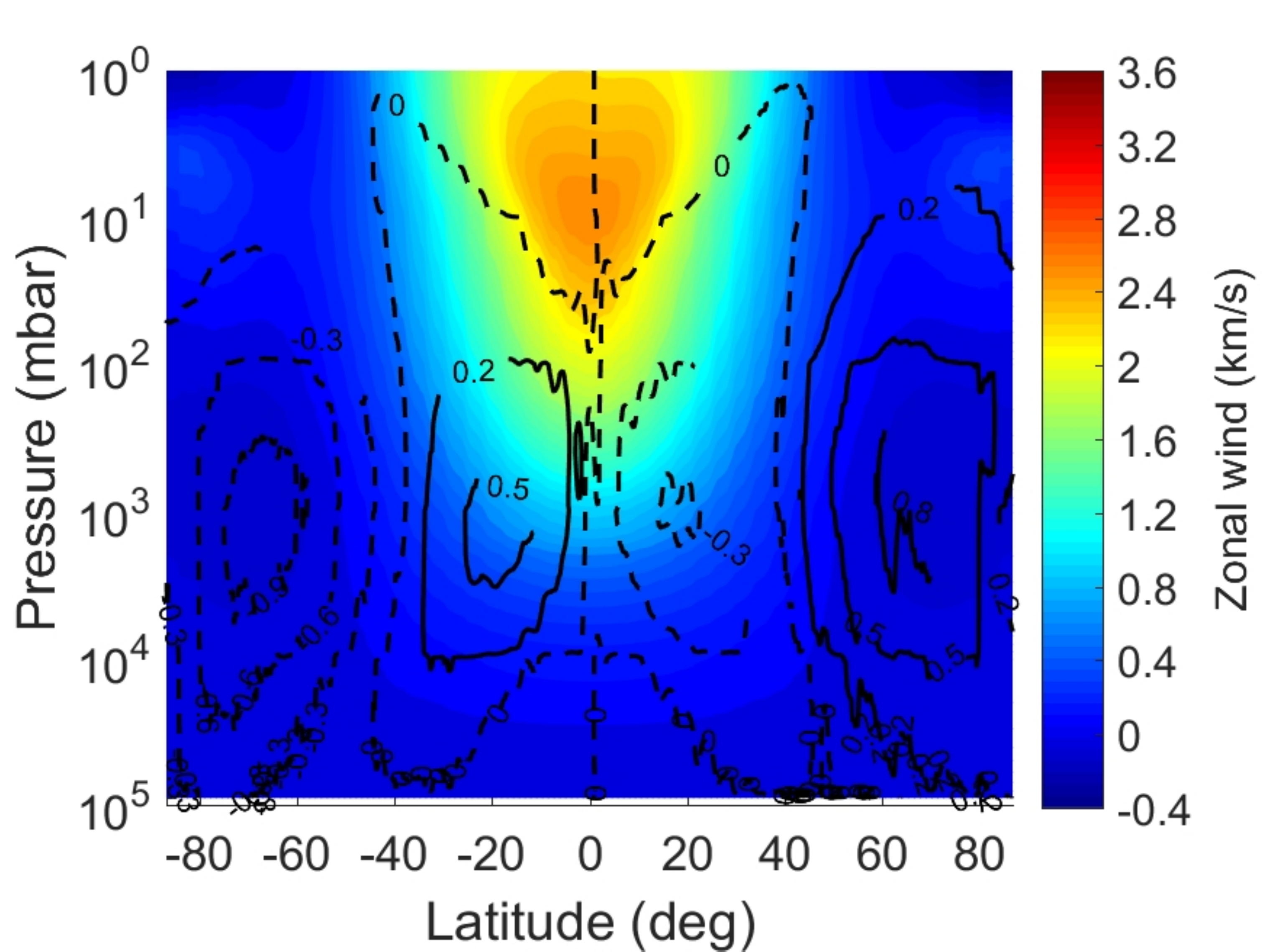}}
\subfigure[Temperature - with clouds]{
\includegraphics[width=0.65\columnwidth]{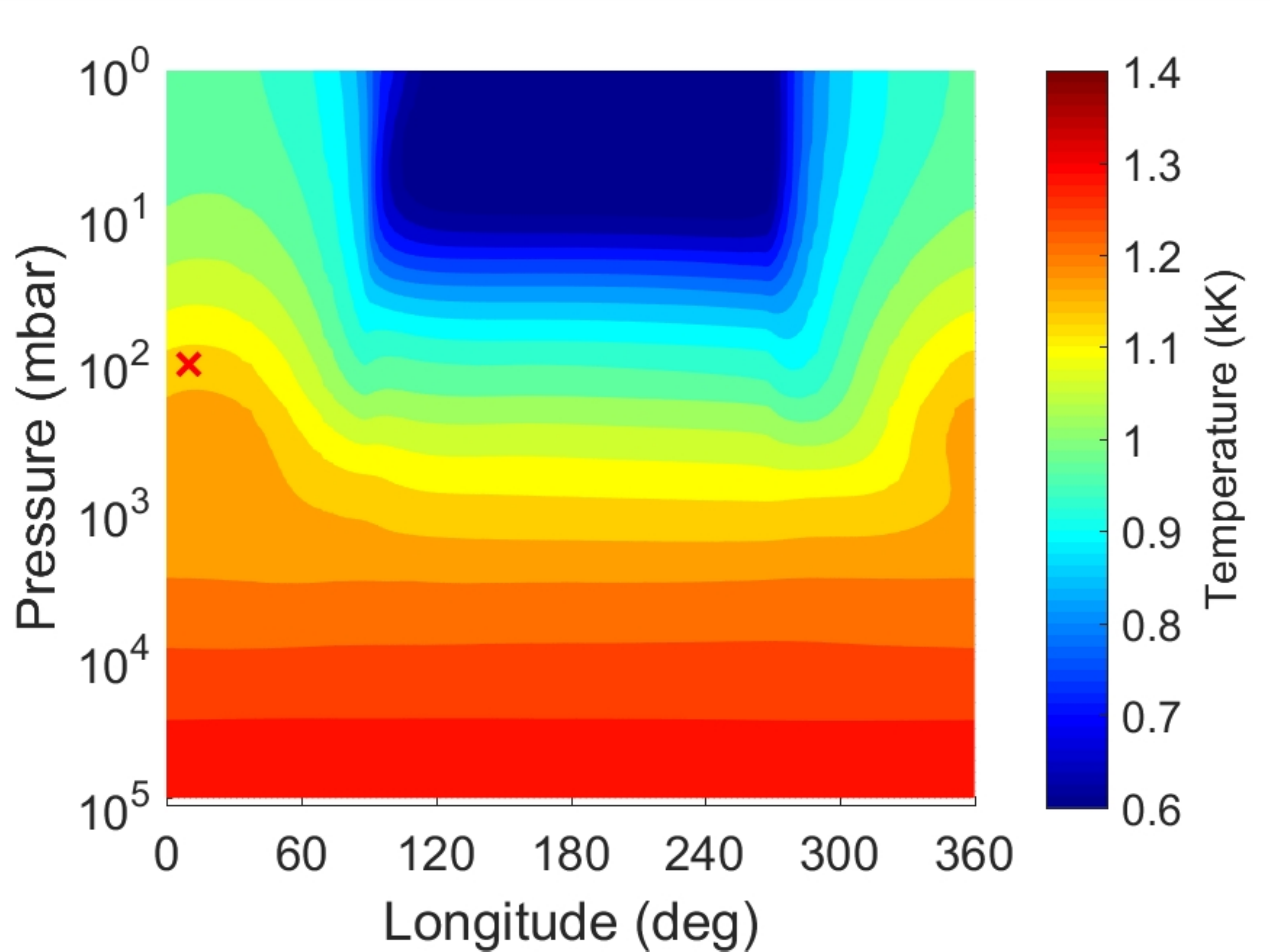}}
\subfigure[Temperature at 10 mbar - with clouds]{
\includegraphics[width=0.65\columnwidth]{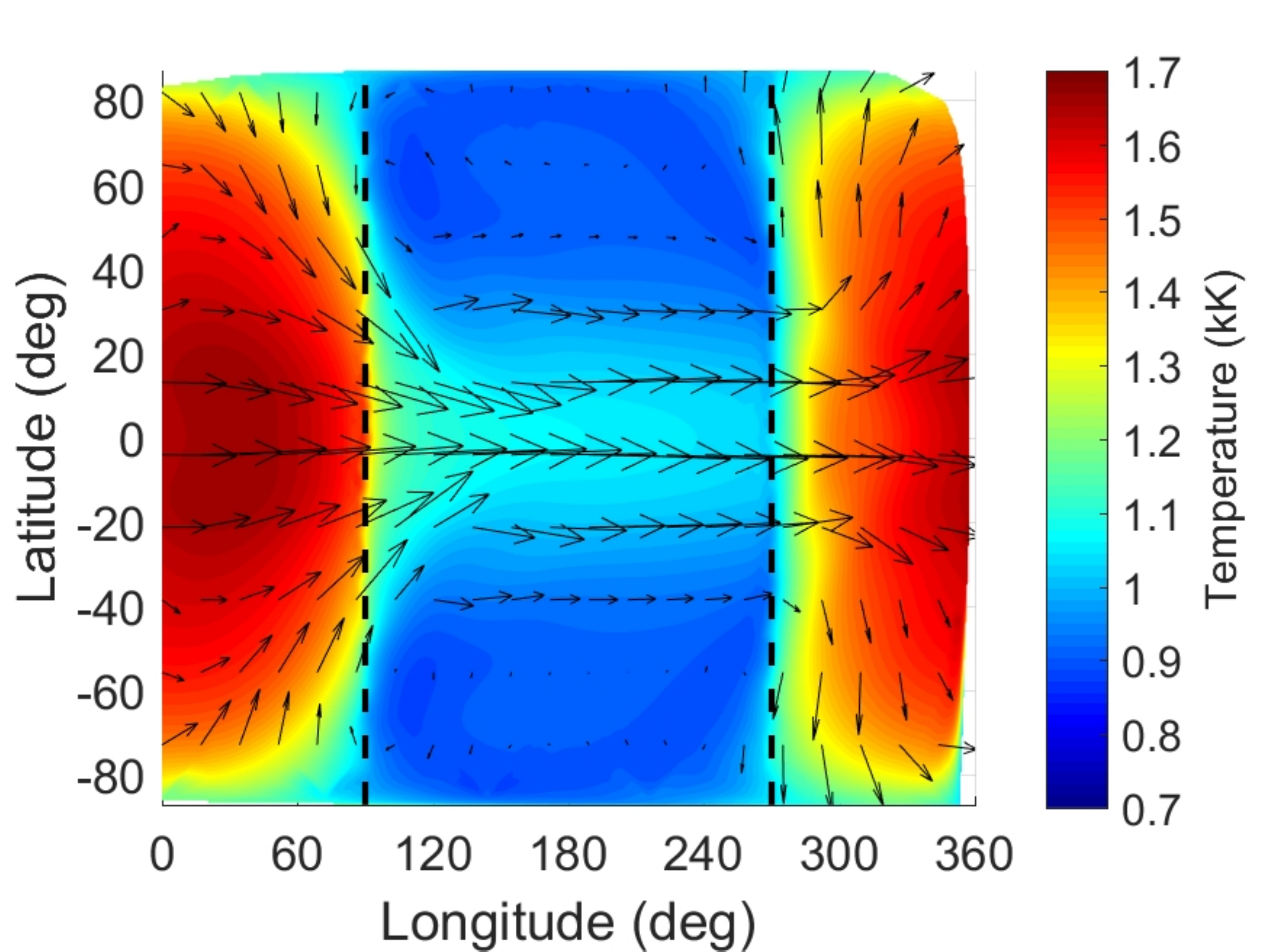}}
\caption{\textbf{(a)} and \textbf{(d)} are the zonal and time averaged zonal winds for the simulations with and without opaque clouds in the nightside of the planet, respectively. The lines are contours of the averaged mass-stream function (in units of 10$^{13}$kg/s). The dashed lines represent the anti-clockwise circulation and the solid lines the clockwise.  \textbf{(b)} and \textbf{(e)} are the maps of temperature averaged in time and latitude. The latitudinal averaging was weighted by the cosine of latitude. \textbf{(c)} and \textbf{(f)} are horizontal maps of temperature at 10 mbar. The arrows shows the time averaged direction of the wind speed. All the results shown in this figure were averaged over the last 500 Earth days of the long simulation. The long-time averaging ensures that the atmospheric structures shown in these plots are not transient features, since the radiative time-scales below the pressure level 1 bar are of the order of hundreds of days (\citealt{2005Iro}). The red crosses in \textbf{(b)} and \textbf{(e)} mark the temperature peaks at 100 mbar and the vertical dashed lines in \textbf{(c)} and \textbf{(f)} the terminators of the planet.}
\label{fig:ref_results_u_temp}
\end{centering}
\end{figure*}

\section{Results}
\label{sec:results}

\subsection{Reference simulations}
\label{sec:ref_simu}

\begin{figure*}
\label{fig:phcv}
\begin{centering}
\subfigure[0.1250 ($\sim$ nightside)]{
\includegraphics[width=0.65\columnwidth]{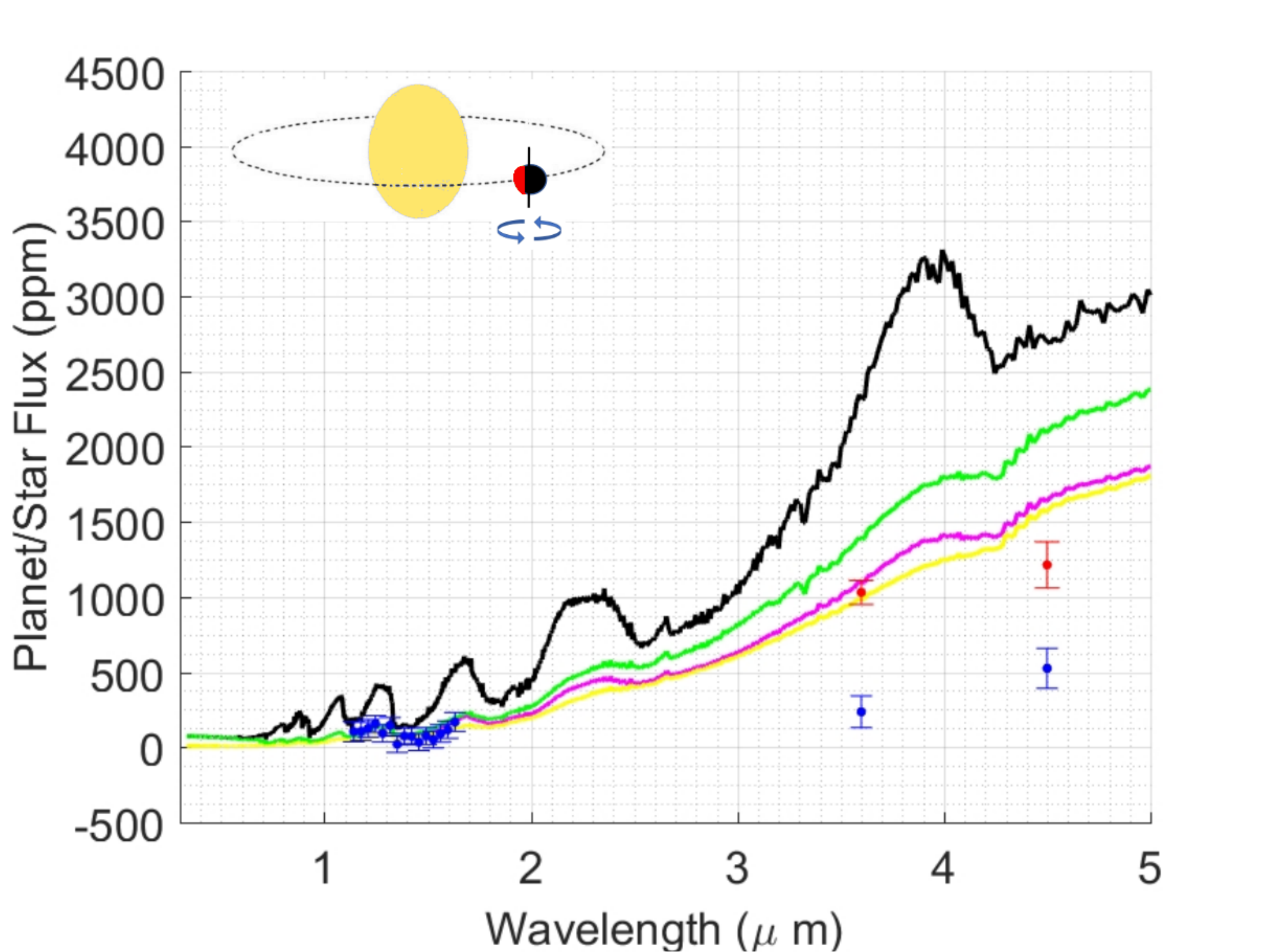}}
\subfigure[0.1875]{
\includegraphics[width=0.65\columnwidth]{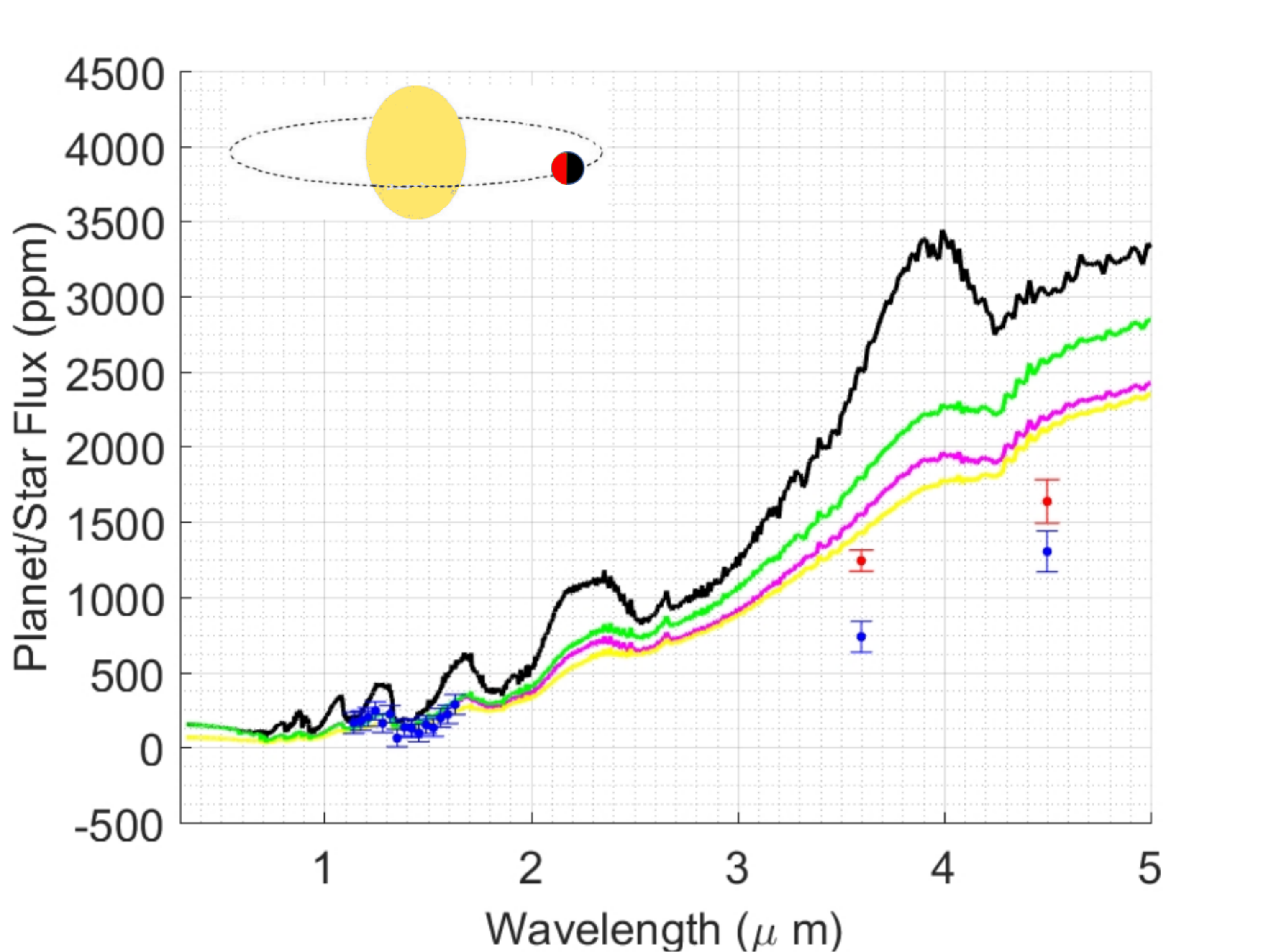}}
\subfigure[0.3125]{
\includegraphics[width=0.65\columnwidth]{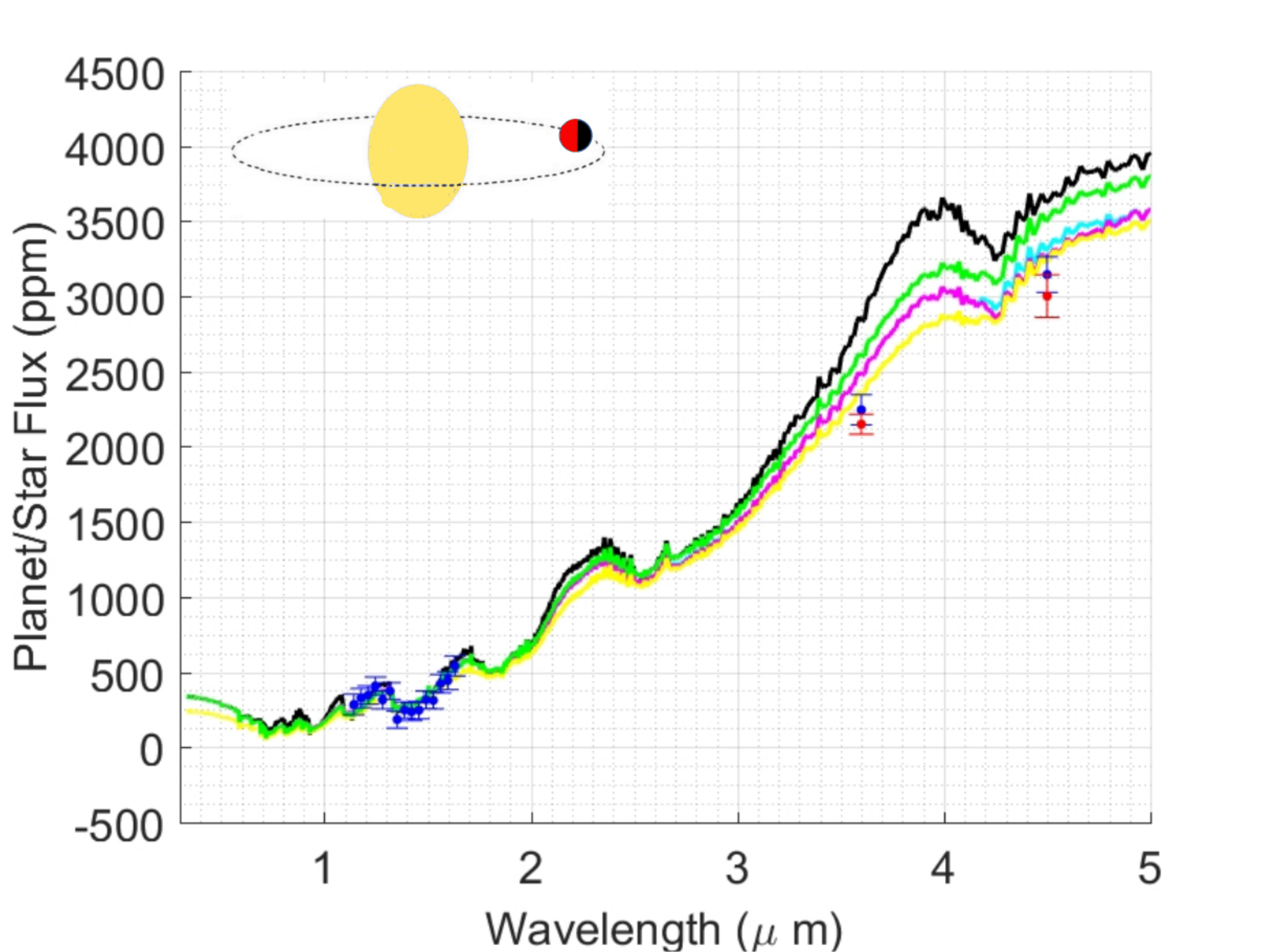}}
\subfigure[0.4375]{
\includegraphics[width=0.65\columnwidth]{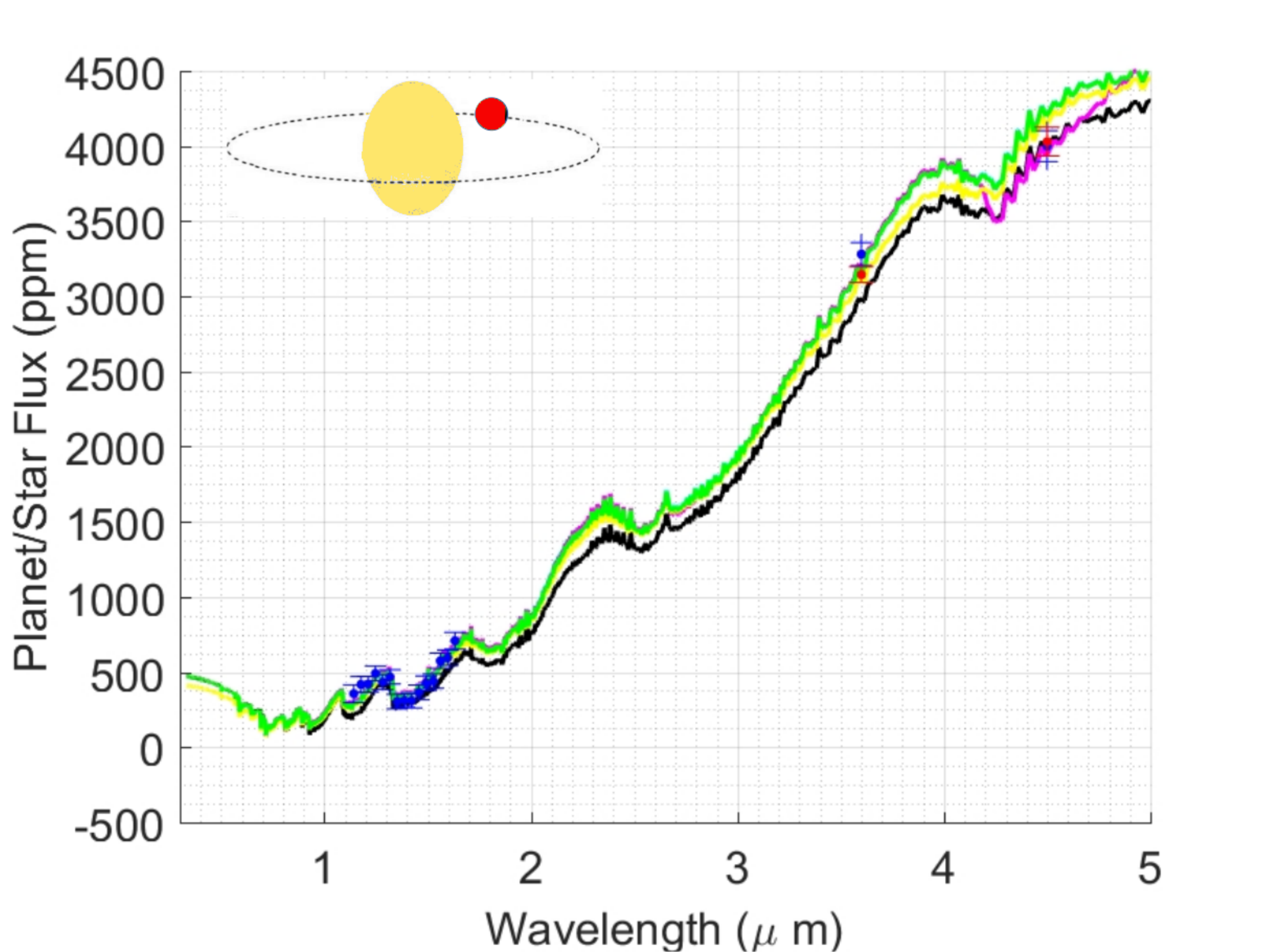}}
\subfigure[0.5000 (dayside)]{
\includegraphics[width=0.65\columnwidth]{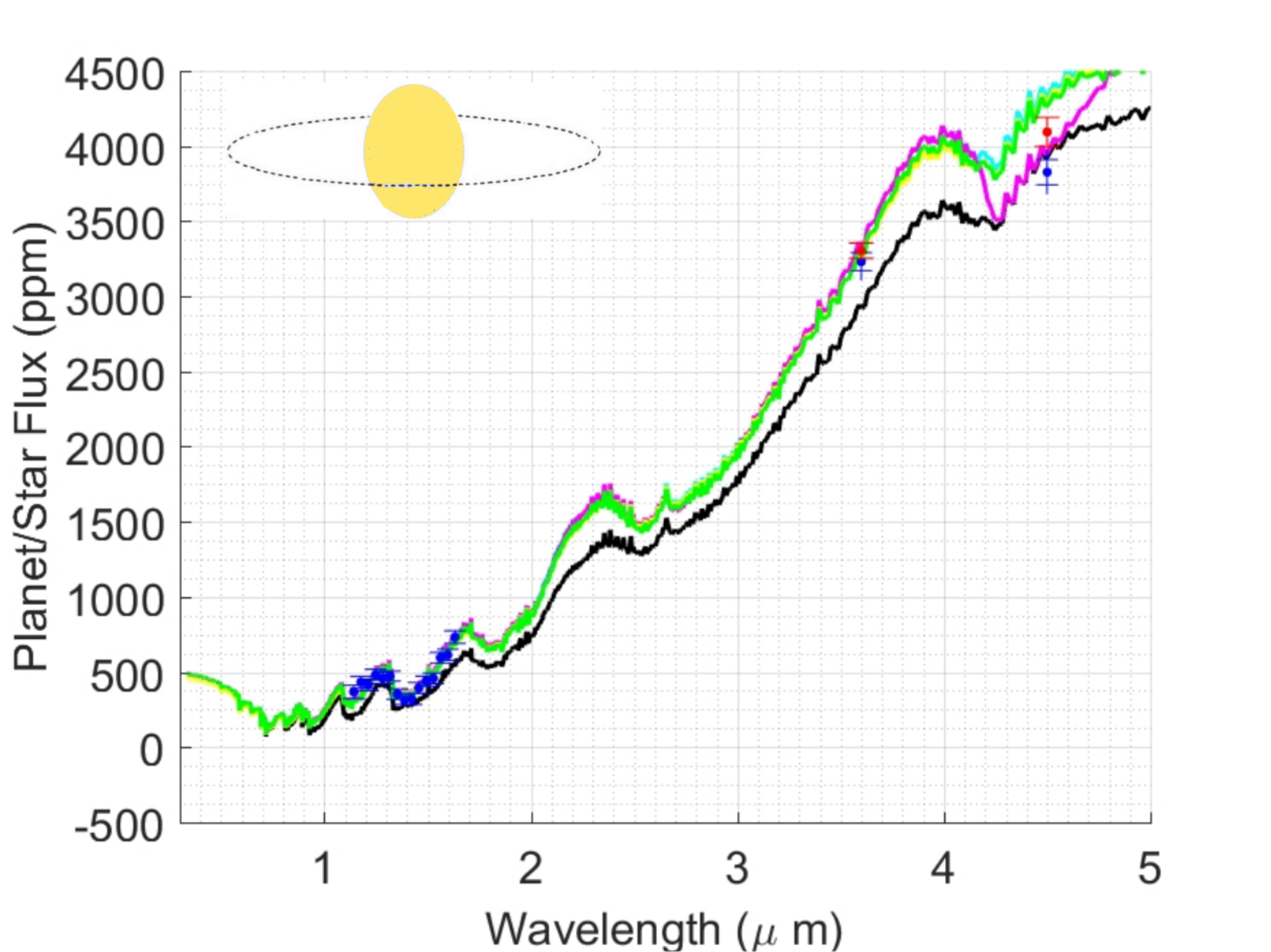}}
\subfigure[0.5625]{
\includegraphics[width=0.65\columnwidth]{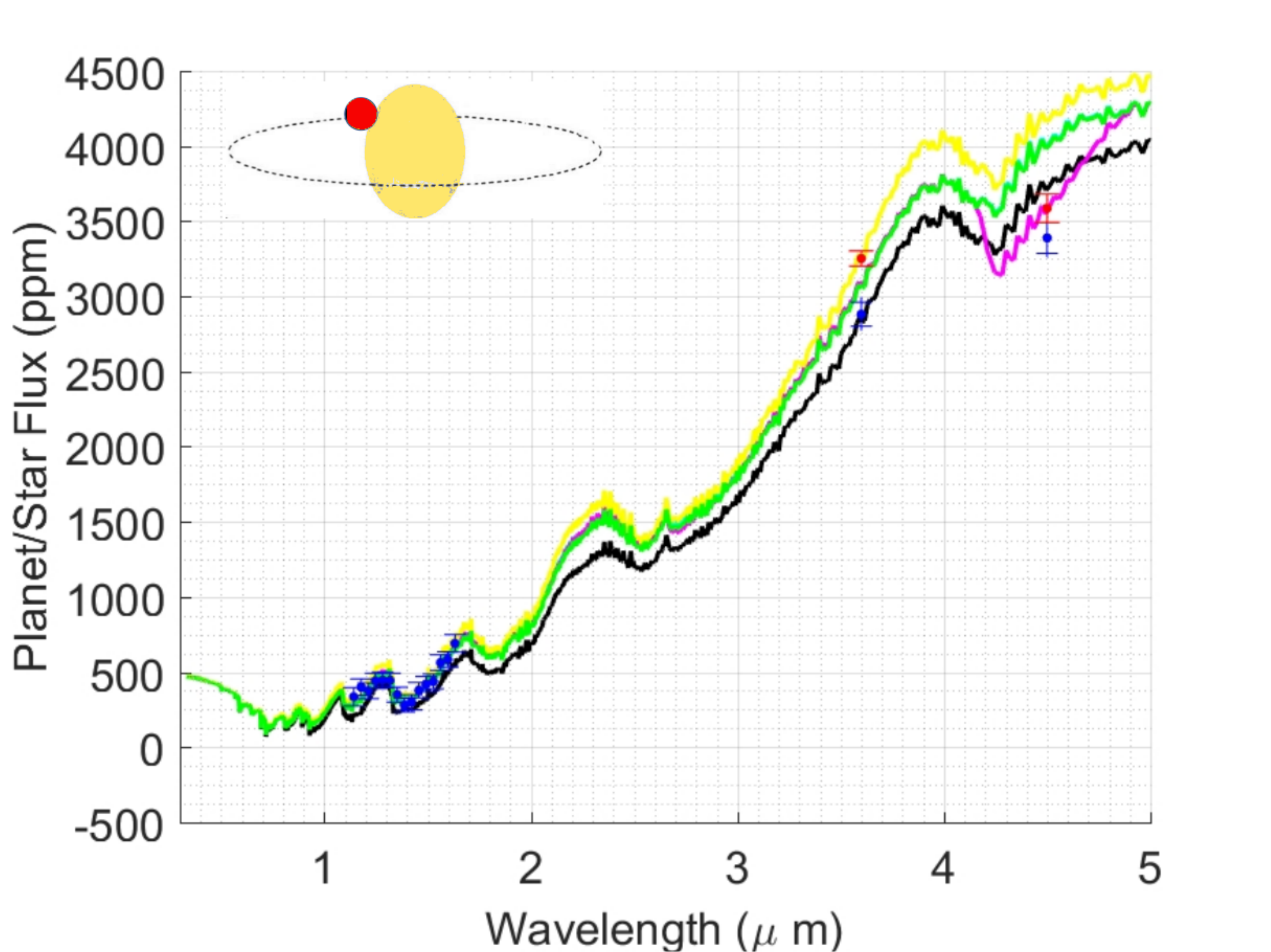}}
\subfigure[0.6875]{
\includegraphics[width=0.65\columnwidth]{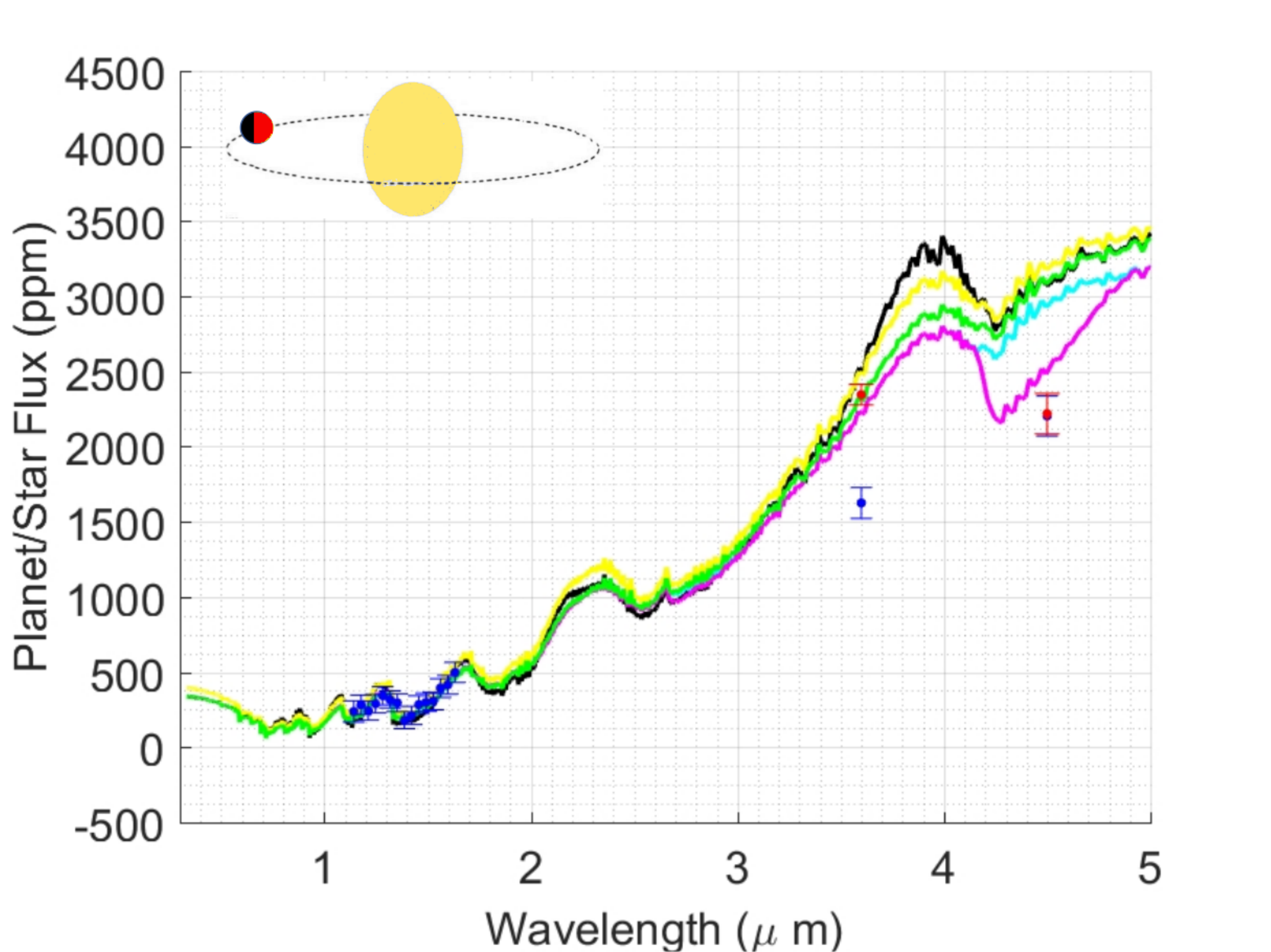}}
\subfigure[0.7500]{
\includegraphics[width=0.65\columnwidth]{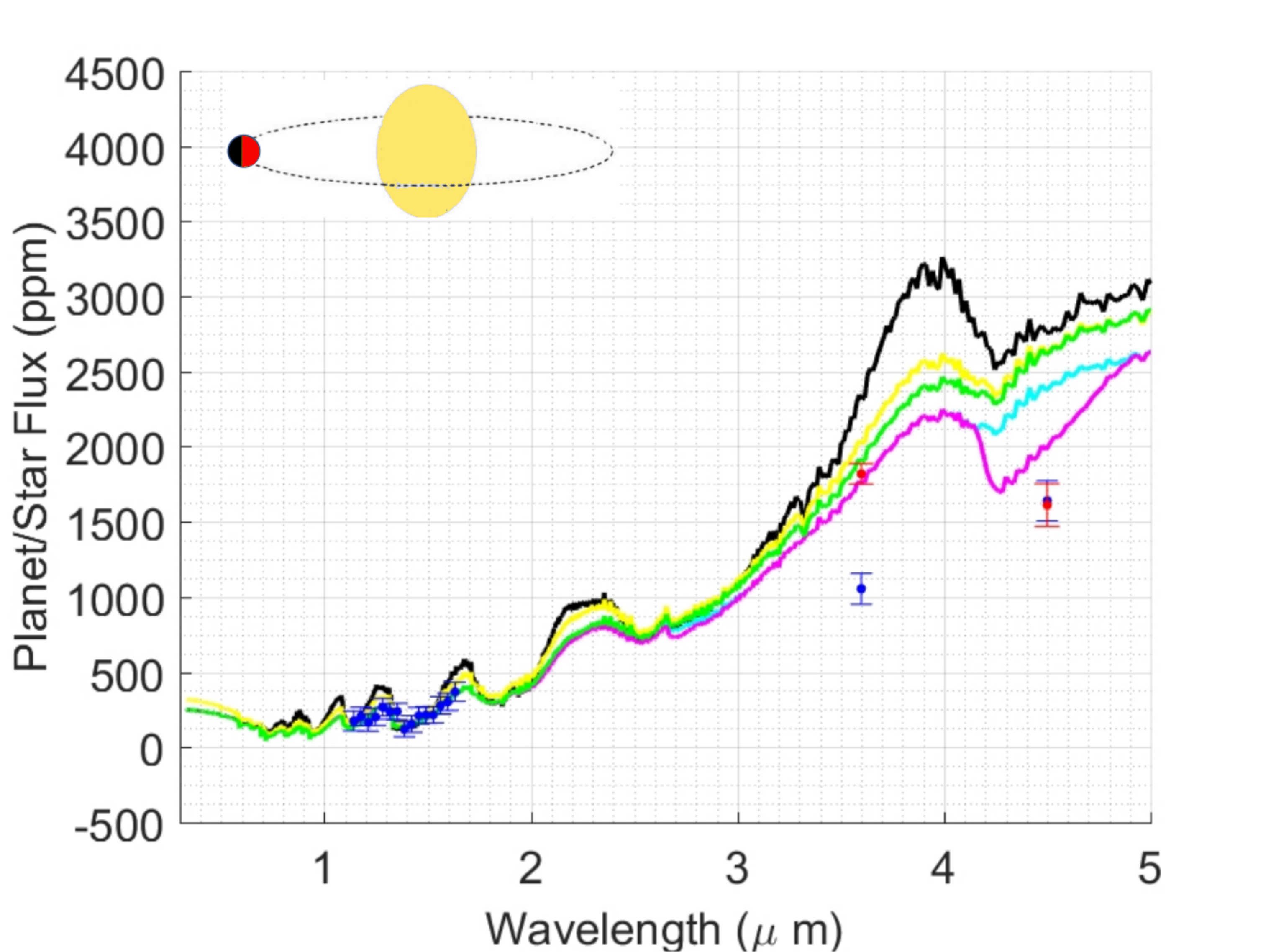}}
\subfigure[0.8750 ($\sim$ nightside)]{
\includegraphics[width=0.65\columnwidth]{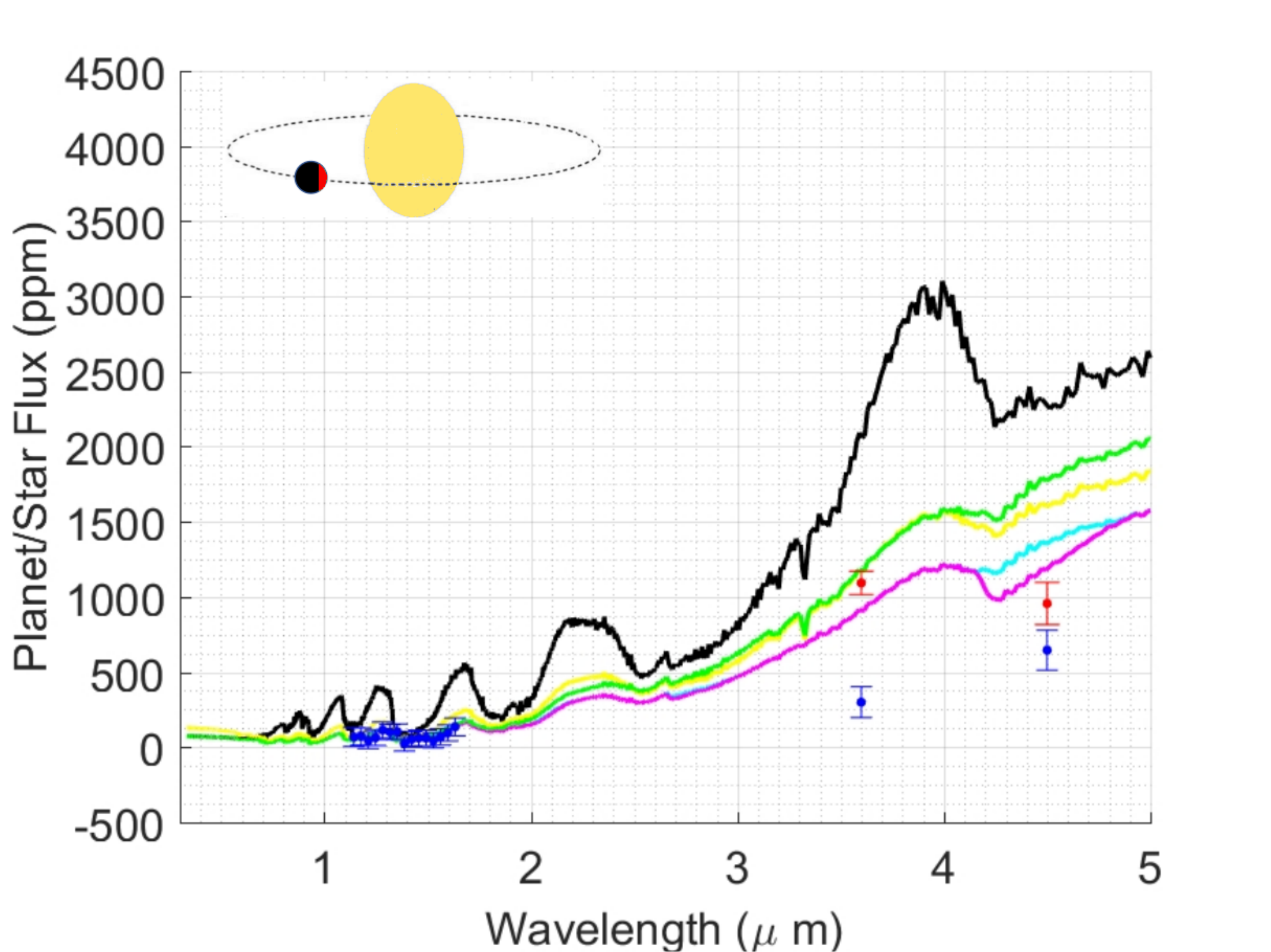}}
\caption{Emission spectra at different orbital phases (panels $a$ to $i$). The primary transit occurs at orbital phase 0.0 and the secondary eclipse at 0.5 (panel $e$). The blue points are \textit{WFC3} data from \cite{2014Stevenson} and \textit{Spitzer} data from \cite{2017Stevenson}. The red points are from our re-analysis of the \textit{Spitzer} data.  The different solid lines correspond to atmospheric scenarios: black - without clouds; cyan - with clouds; magenta - with clouds and extra CO$_2$; yellow - clouds in the night side shifted 20 degrees westwards; green - clouds with lower cloud top level (20 mbar instead of 10 mbar).}
\label{fig:planet_spec}
\end{centering}
\end{figure*}

We present two GCM runs in Figure \ref{fig:ref_results_u_temp}, where the top and bottom rows show the cloudfree and cloudy GCMs, respectively.  The output from these GCMs will be post-processed to produce multi-phase emission spectra and multi-wavelength phase curves, which we will discuss shortly.  Here, we point out the lessons learned from comparing the cloudfree and cloudy GCMs.  First, both GCMs show the presence of an equatorial zonal jet, which is ubiquitous in all current GCMs of hot Jupiters (see \citealt{2015Heng} for a review). The main mechanism driving the formation of the equatorial jet is the equator-ward transport of angular momentum (i.e., an upgradient transport of angular momentum) caused by a tilt of the diurnal tide phase front with respect to the latitude, in the latitude-longitude plane (e.g., \citealt{2011Showman,2014Tsai}). The presence of clouds on the nightside weakens the equatorial zonal jet in our simulations, as the mechanism responsible for the transport of angular momentum towards the equator is partially disrupted.  The zonally averaged mass streamfunction profiles for both GCMs reveal the presence of anti-Hadley circulation cells, which transport mass and heat downwards (to greater pressures) at the equator.  Second, the presence of nightside clouds results in a greater temperature contrast between the dayside and nightside of WASP-43b.  Corresponding, the shift of the peak of the thermal phase curves, which track the temperature profile across longitude, is reduced by the presence of clouds.  Third, the chevron feature at 10 mbar is robust to the presence of clouds, but is altered somewhat by the cloud-driven change in the atmospheric circulation. Of interest are the large-scale vortices at mid-latitudes, which are the coldest regions of the atmosphere at 10 mbar. These cold regions can be associated with regions that can trap and grow larger cloud particles (see for example \citealt{2016Lee}).  Overall, we expect the influence of the clouds on the variation of temperature across altitude, latitude and longitude to be manifested in the emission spectra and phase curves, which we will now explore.

\subsection{Different physical scenarios explored}
\label{sec:suite}

To understand the influence of clouds in the atmosphere of WASP-43b, we explore the following idealized scenarios.
\begin{itemize}

\item A cloudfree atmosphere, shown by the black curves in both Figures \ref{fig:planet_spec} and \ref{fig:phase-curves}. This experiment is consistent with the work from \cite{2015Kataria}.

\item A cloudy atmosphere with a cloud deck on the nightside extending to a cloud-top pressure of 10 mbar (cyan curves).  This mimic a cloud deck forming due to the cooler temperatures of the nightside as suggested by \cite{2015Kataria}.

\item A cloudy atmosphere with a nightside cloud deck and enhanced carbon dioxide (CO$_2$) (magenta curves).  The consideration of CO$_2$ is motivated by its ability to absorb radiation at 4.5 $\mu$m relative to other molecules.  To mimic its change in abundance from the dayside to the nightside of WASP-43b, we assume additional CO$_2$ to be absent at the substellar point (noon) and increase to a mass mixing ratio of $10^{-3}$ at the antistellar point.  We emphasize that this is in addition to CO$_2$ that is assumed to be present according to chemical equilibrium and local conditions of temperature and pressure. This scenario represents a similar chemical disequilibrium process driven by the atmospheric transport as suggested by \cite{2005Cooper} and \cite{2014Agundez}.

\item A cloudy atmosphere with the cloud deck shifted westwards in longitude by 20 degrees (yellow curves).  Physically, it mimics the protrusion of the cloud deck from the nightside into the dayside caused by atmospheric circulation and the presence of the cold vortices at mid-latitude. This experiment could represent a scenario similar to Kepler-7b \citep{2013Demory}.

\item A cloudy atmosphere with the cloud top located at 20 mbar instead of 10 mbar (green curves).  This mimics variation in the microphysical cloud processes and atmospheric mixing that we are not modeling from first principles (e.g., \citealt{2016Parmentier}).

\end{itemize}

\subsection{Constraints from multi-phase emission spectra}
\label{subsec:mespectra}

Figure \ref{fig:planet_spec} shows the multi-phase emission spectra corresponding to the suite of models described in \S\ref{sec:suite}, as well as the \textit{Spitzer} data points from both \cite{2017Stevenson} and our re-analysis of the same data.  The first thing to notice is how our re-analysis of the \textit{Spitzer} data has little effect on the dayside emission spectrum, as well as the spectra just before and after the dayside.  However, it has a significant effect on the emission spectra at orbital phases just before and after that of the nightside.  Specifically, the nightside of WASP-43b is now emitting more flux.

Consistent with \cite{2015Kataria}, our cloudfree GCM produces a decent match to the dayside emission spectrum of WASP-43b, as do the GCMs with nightside clouds.  We lay claim to the same statement made by \cite{2015Kataria}, which is that this agreement between model and data is accomplished with little to no finetuning.  

The nightside emission spectra (orbital phases of 0.125 and 0.875), as well as those at orbital phases of 0.1875, 0.3125 and 0.75, are inconsistent with the predictions from our cloudfree GCM, which over-predict both the \textit{WFC3} and \textit{Spitzer} fluxes.  Our re-analysis of the \textit{Spitzer} data points brings them into closer agreement with the models, but it is clear that a cloudy nightside is needed to match the data.  Additionally, the emission spectra at orbital phases of 0.125 and 0.1875 disfavor the scenario in which the cloud top is located at 20 mbar, which produce model 3.6 and 4.5 $\mu$m fluxes that are higher than the measured fluxes.  

At orbital phases of 0.6875, 0.75 and 0.875, the 4.5 $\mu$m \textit{Spitzer} flux is lower than predicted by the cloudy models, unless enhanced CO$_2$ is present in the atmosphere. CO has also an important absorption feature at this wavelength, however, tests assuming an atmosphere composed of 100$\%$ CO could not reduce the fluxes down to the values observed. However, CO will contribute to reduce the fluxes at 4.5 $\mu$m at larger orbital phases, because it is expected that its abundance is enhanced relative to its chemical equilibrium values in the nightside due to the efficient transport from the dayside  (e.g., \citealt{2005Cooper}; \citealt{2014Agundez}). Future simulations capable of computing self-consistently chemistry, radiation and atmospheric dynamics, will be able to elucidate us on the nature of this atmospheric process. There appears to be no evidence in favor of a cloud deck that is shifted westwards in longitude.

\subsection{Constraints from multi-wavelength phase curve}
\label{sec:const-curve}
\begin{figure*}
\label{fig:curves}
\begin{centering}
\subfigure[1.21$\mu$m]{
\includegraphics[width=0.65\columnwidth]{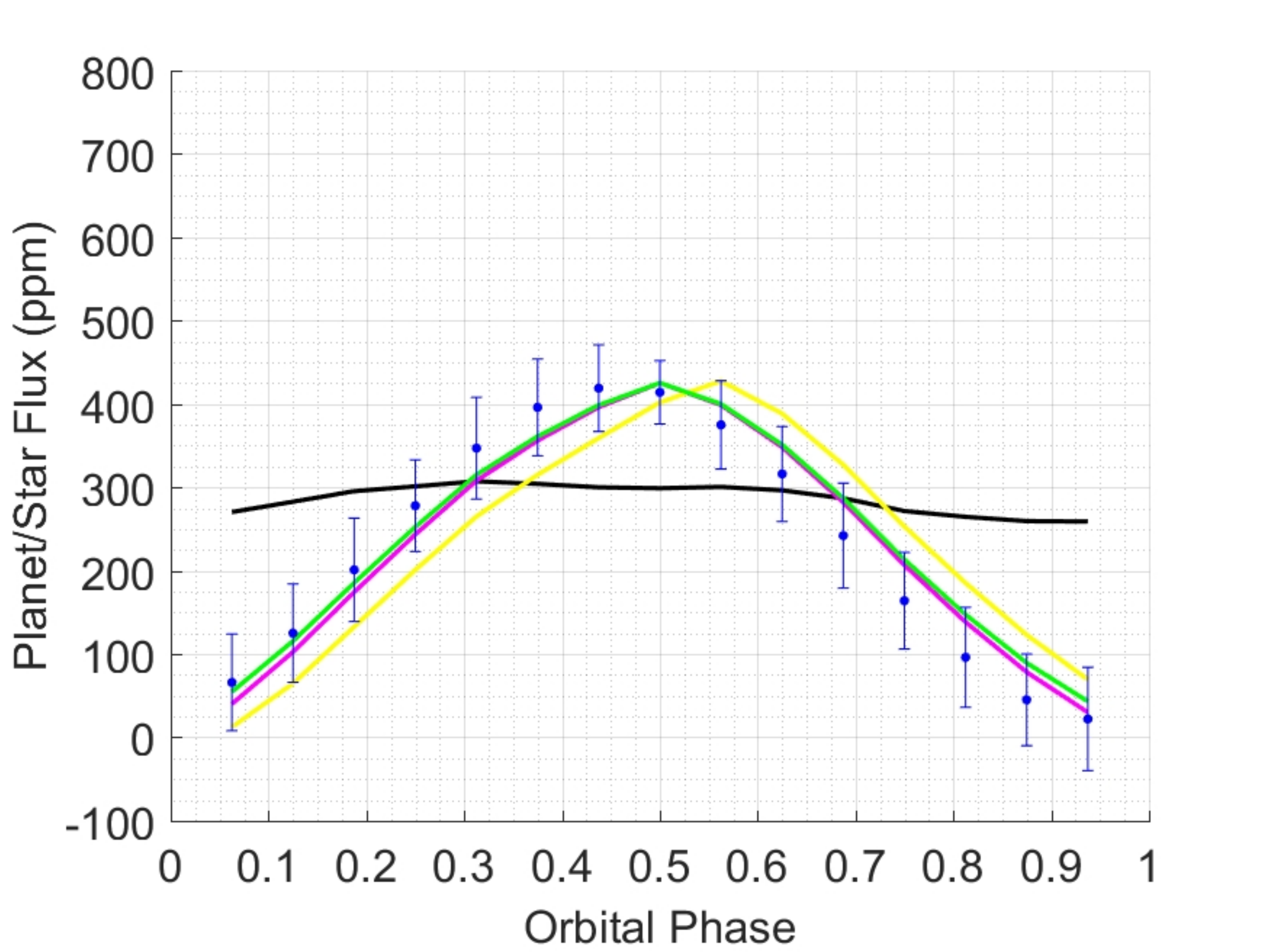}}
\subfigure[1.28$\mu$m]{
\includegraphics[width=0.65\columnwidth]{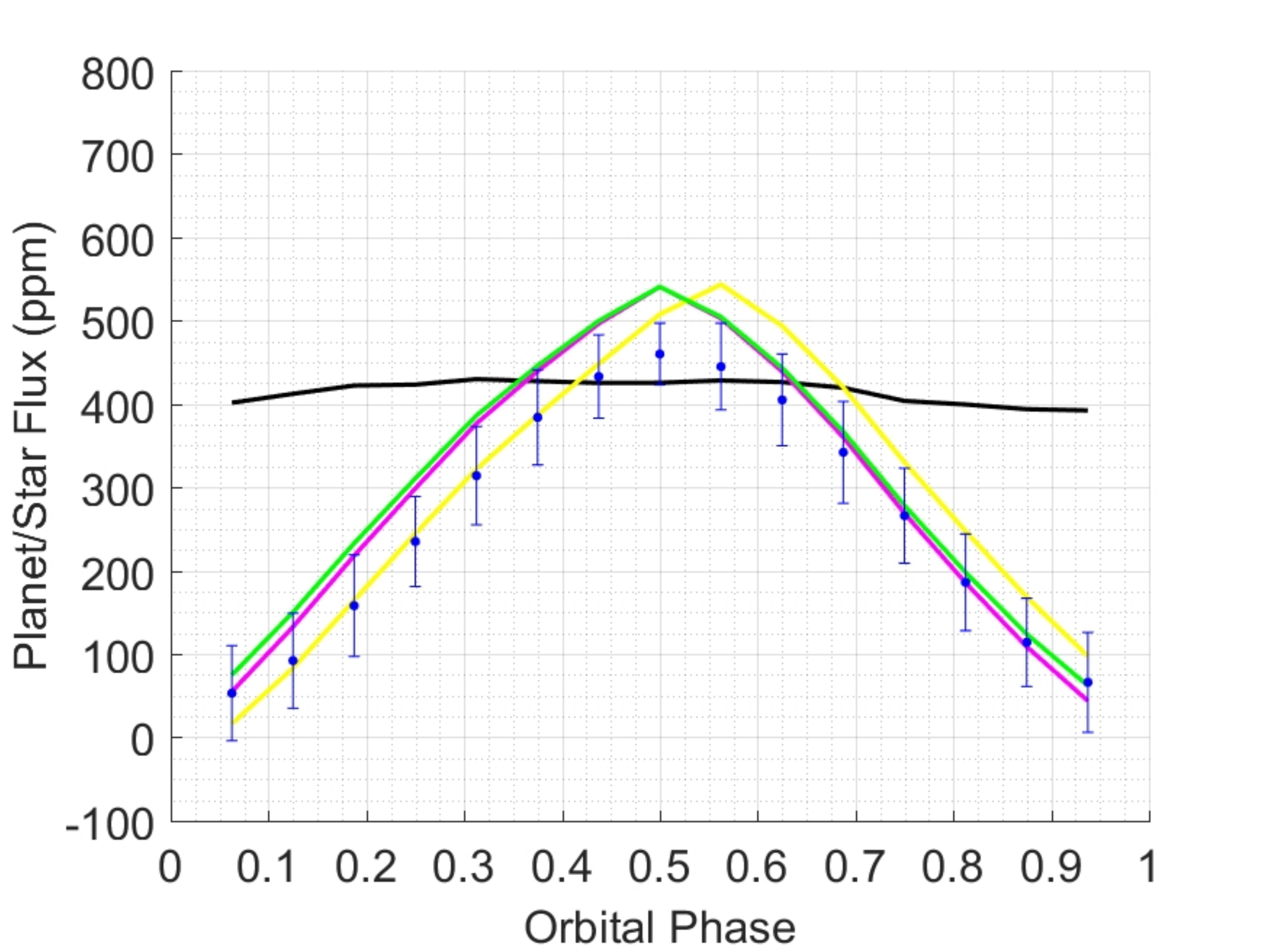}}
\subfigure[1.35$\mu$m]{
\includegraphics[width=0.65\columnwidth]{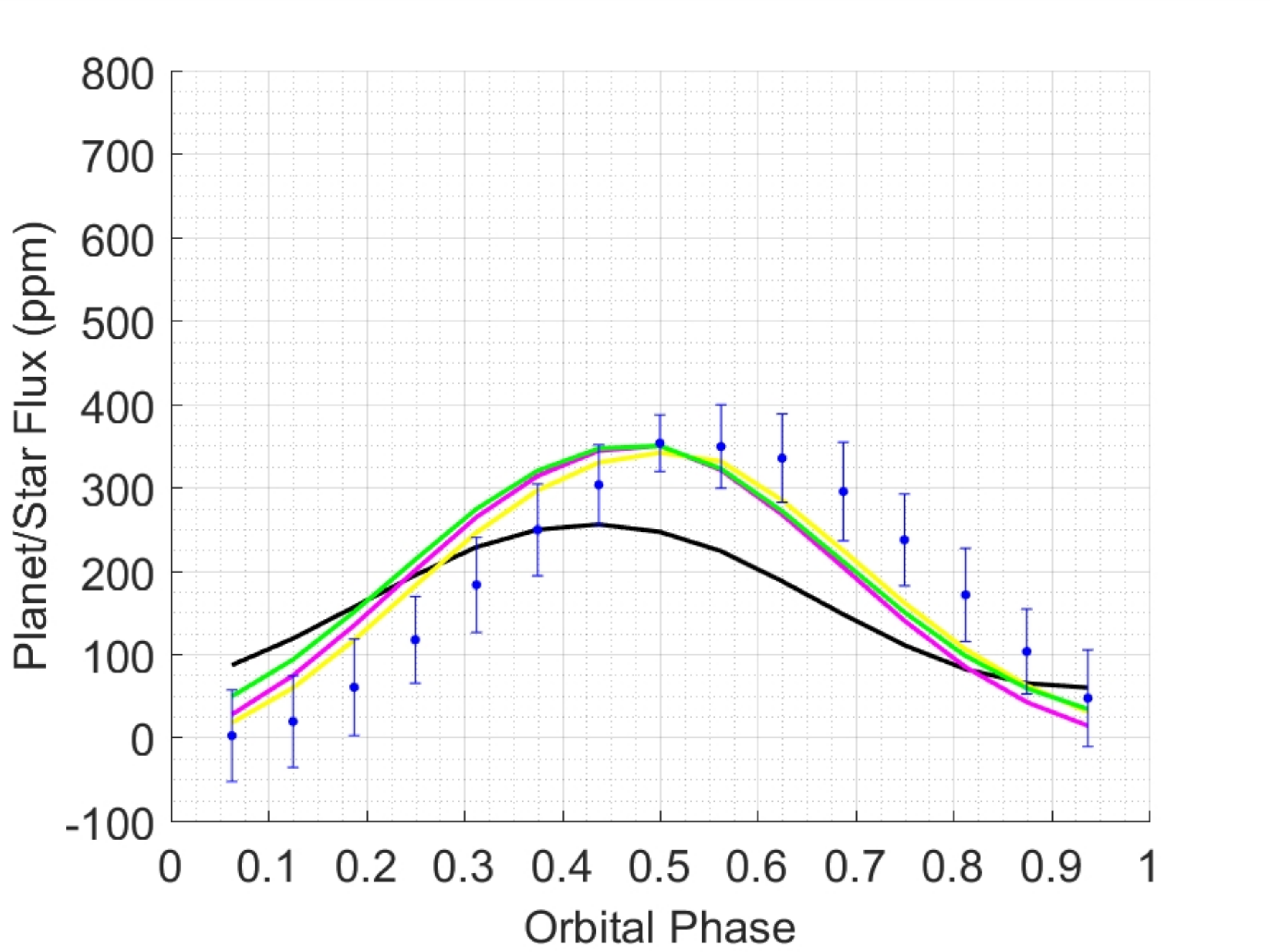}}
\subfigure[1.42$\mu$m]{
\includegraphics[width=0.65\columnwidth]{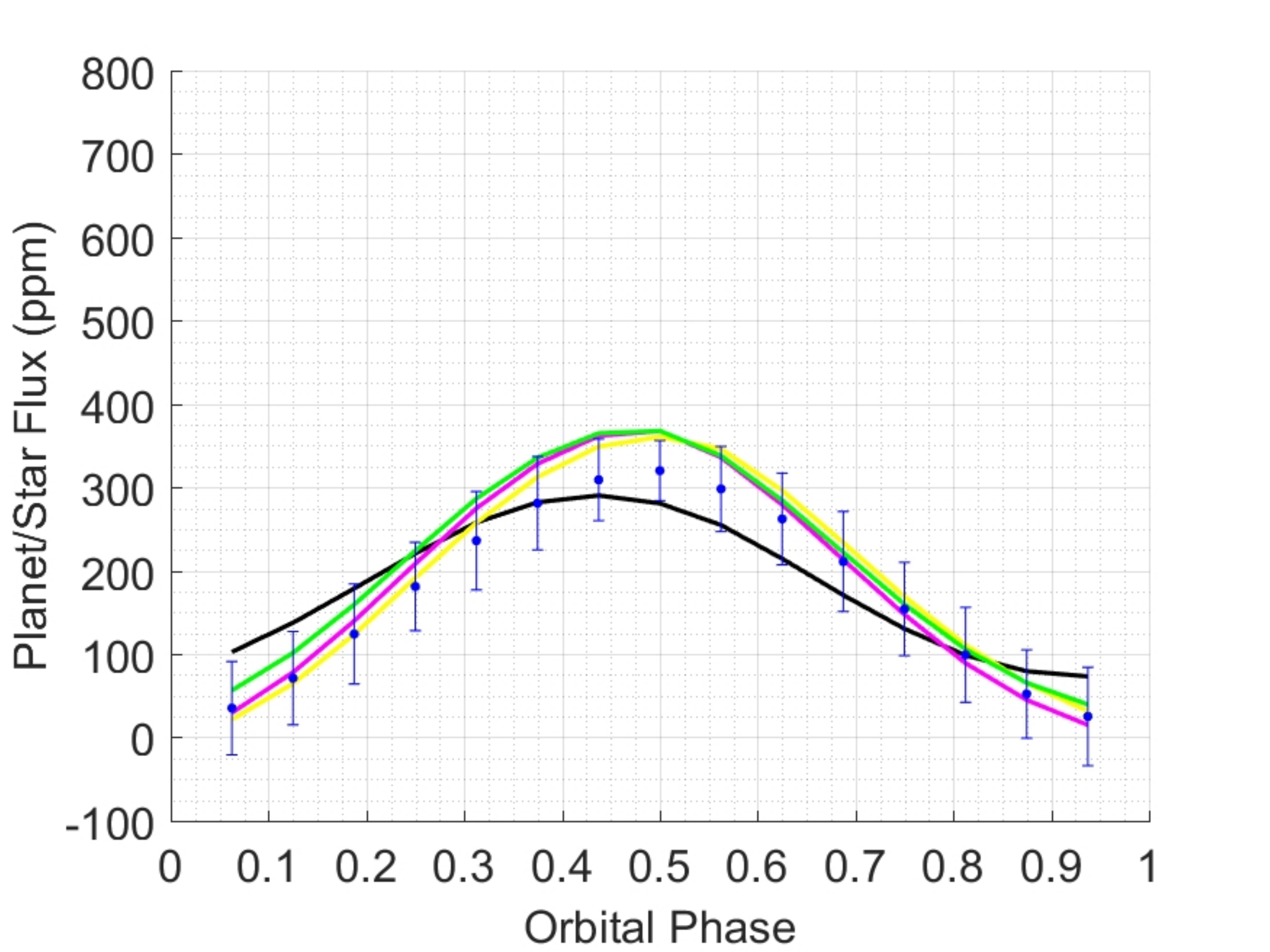}}
\subfigure[1.49$\mu$m]{
\includegraphics[width=0.65\columnwidth]{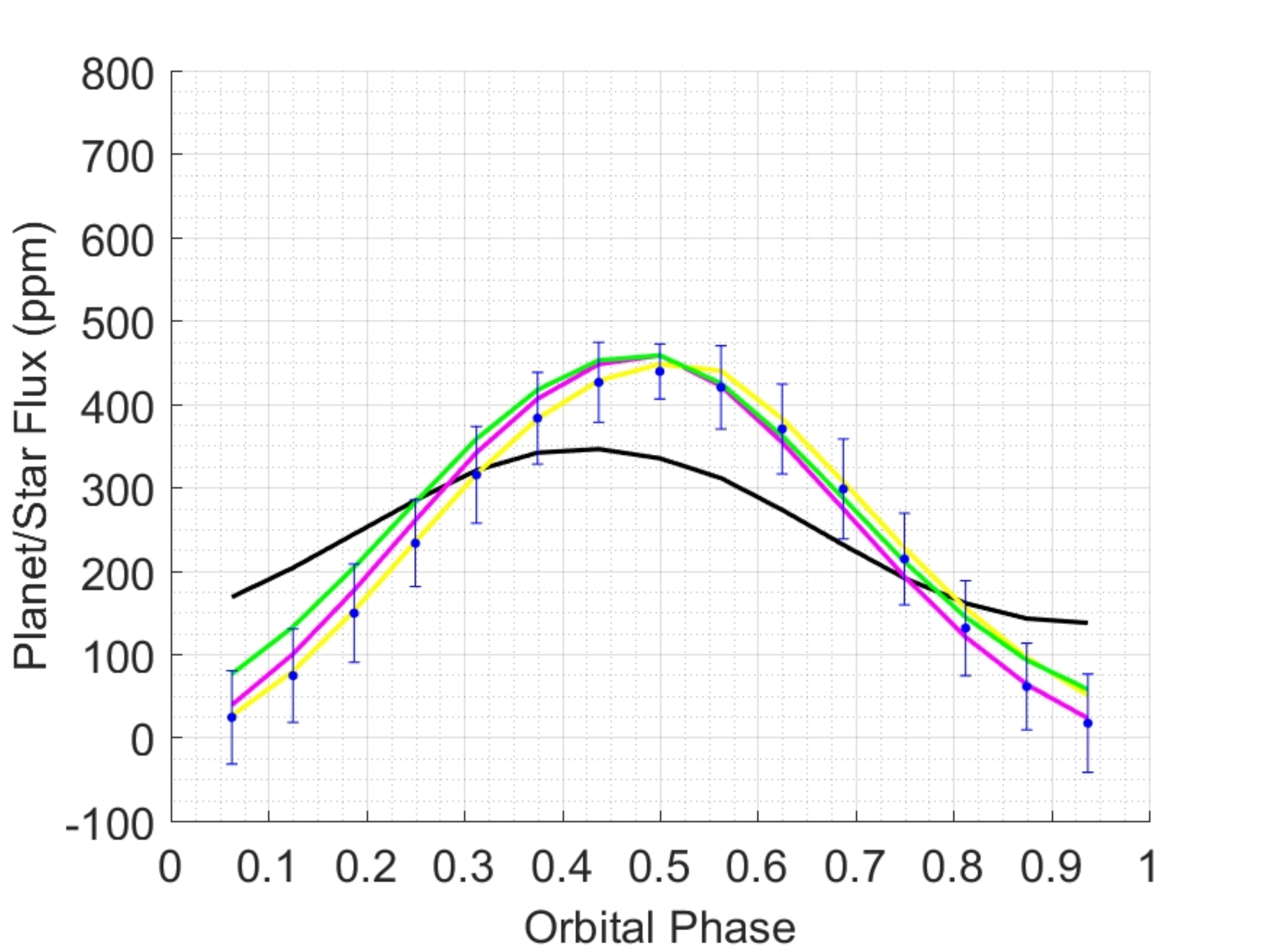}}
\subfigure[1.56$\mu$m]{
\includegraphics[width=0.65\columnwidth]{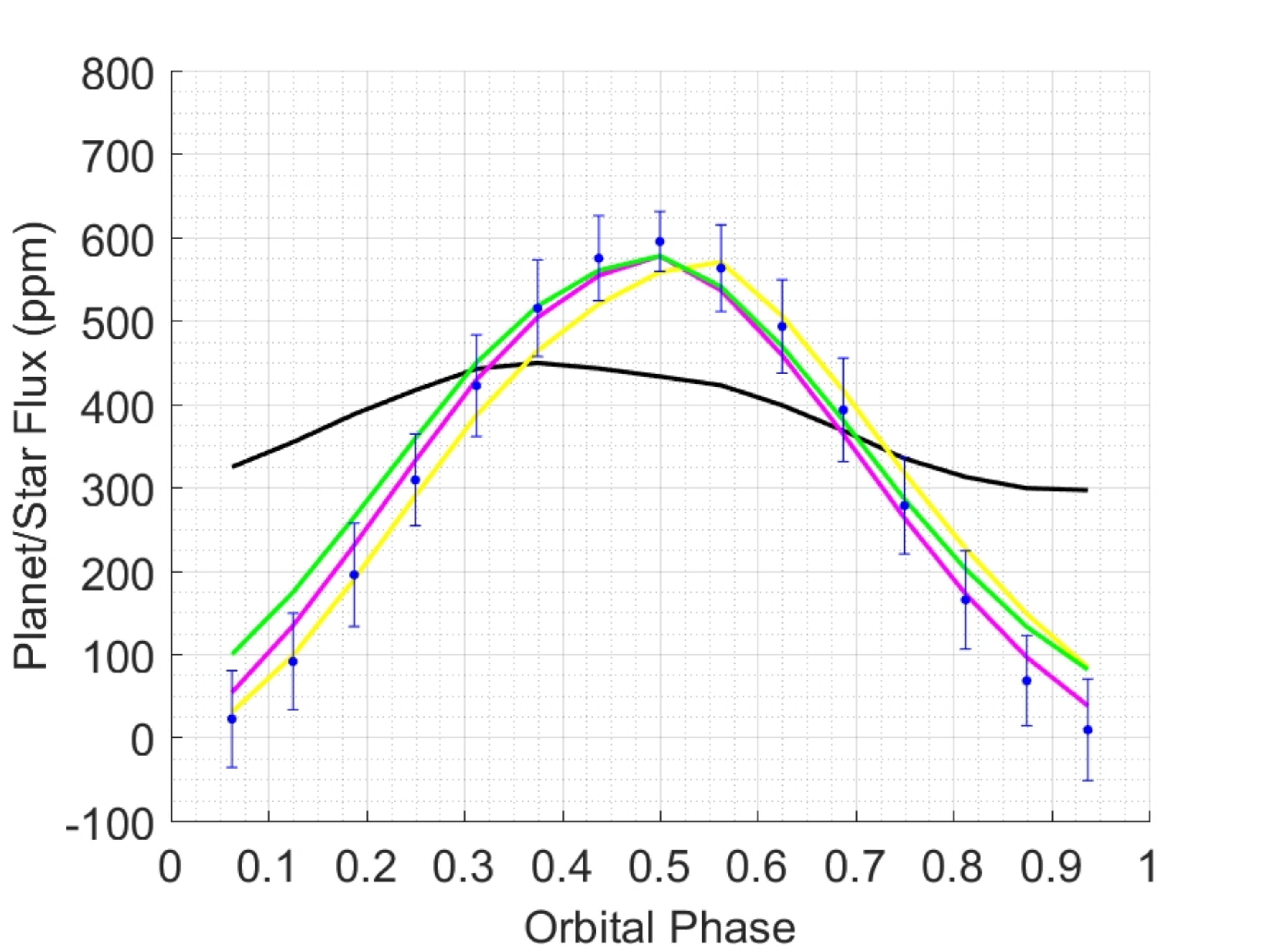}}
\subfigure[1.63$\mu$m]{
\includegraphics[width=0.65\columnwidth]{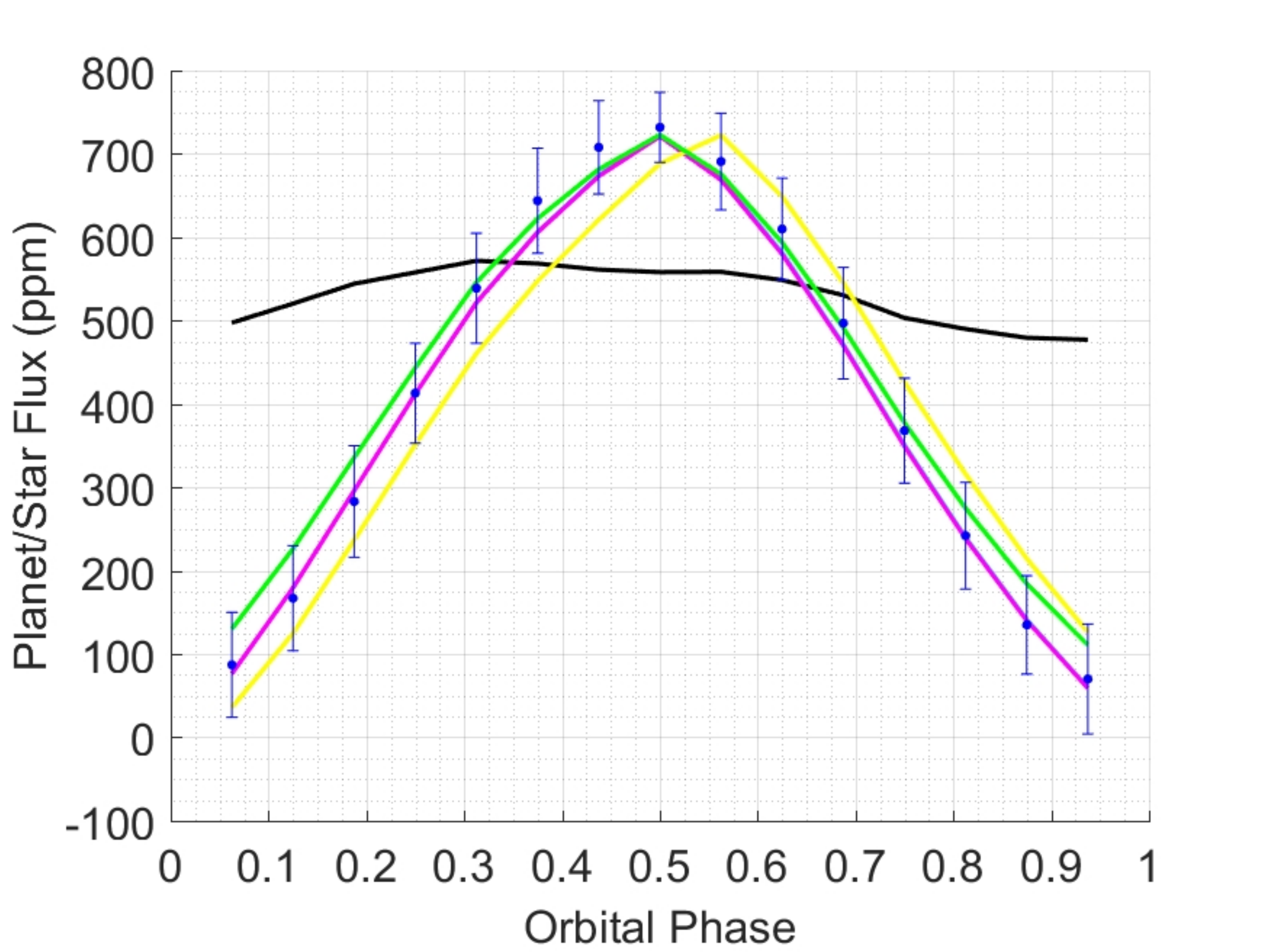}}
\subfigure[3.6$\mu$m]{
\includegraphics[width=0.65\columnwidth]{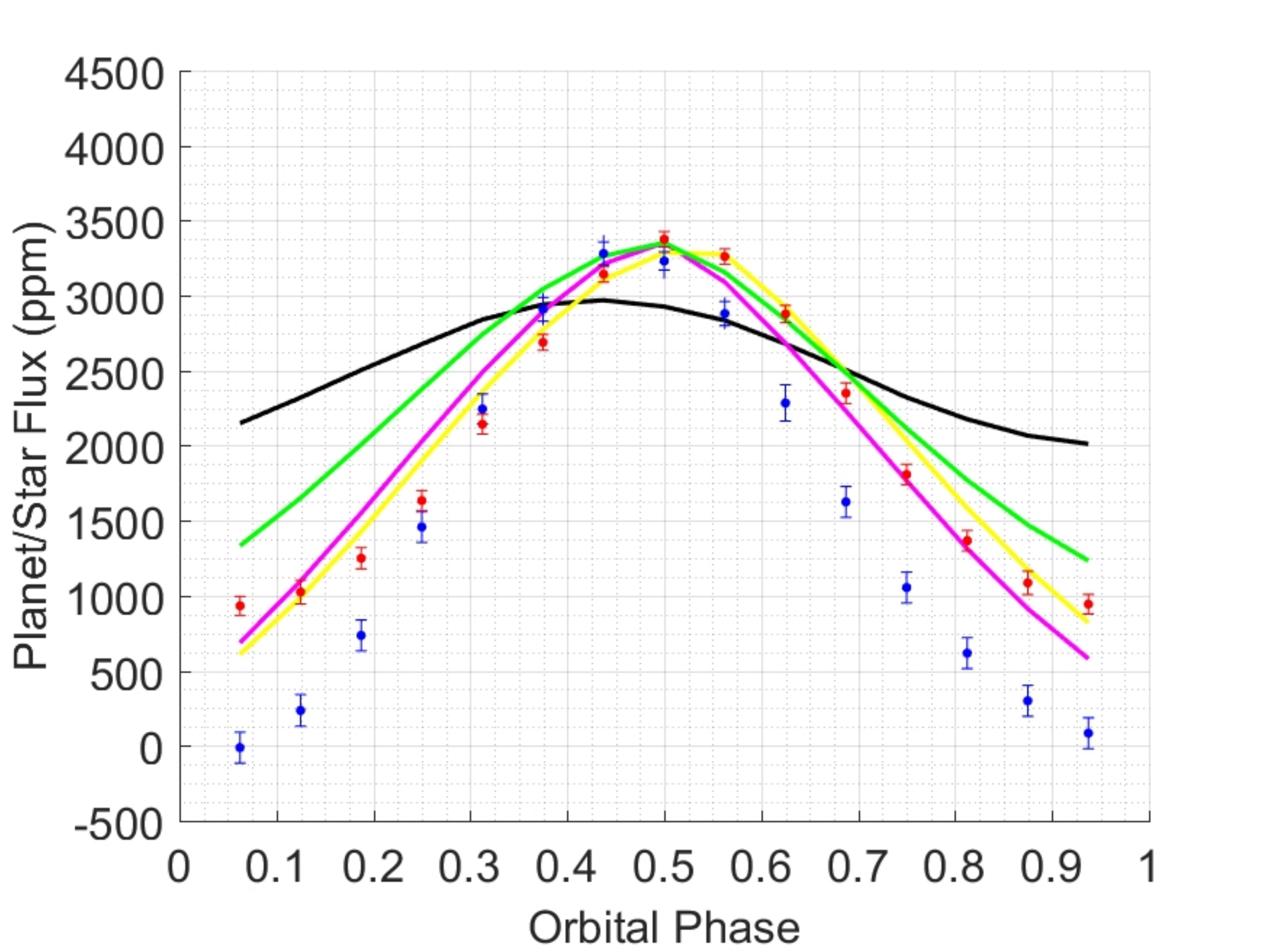}}
\subfigure[4.5$\mu$m]{
\includegraphics[width=0.65\columnwidth]{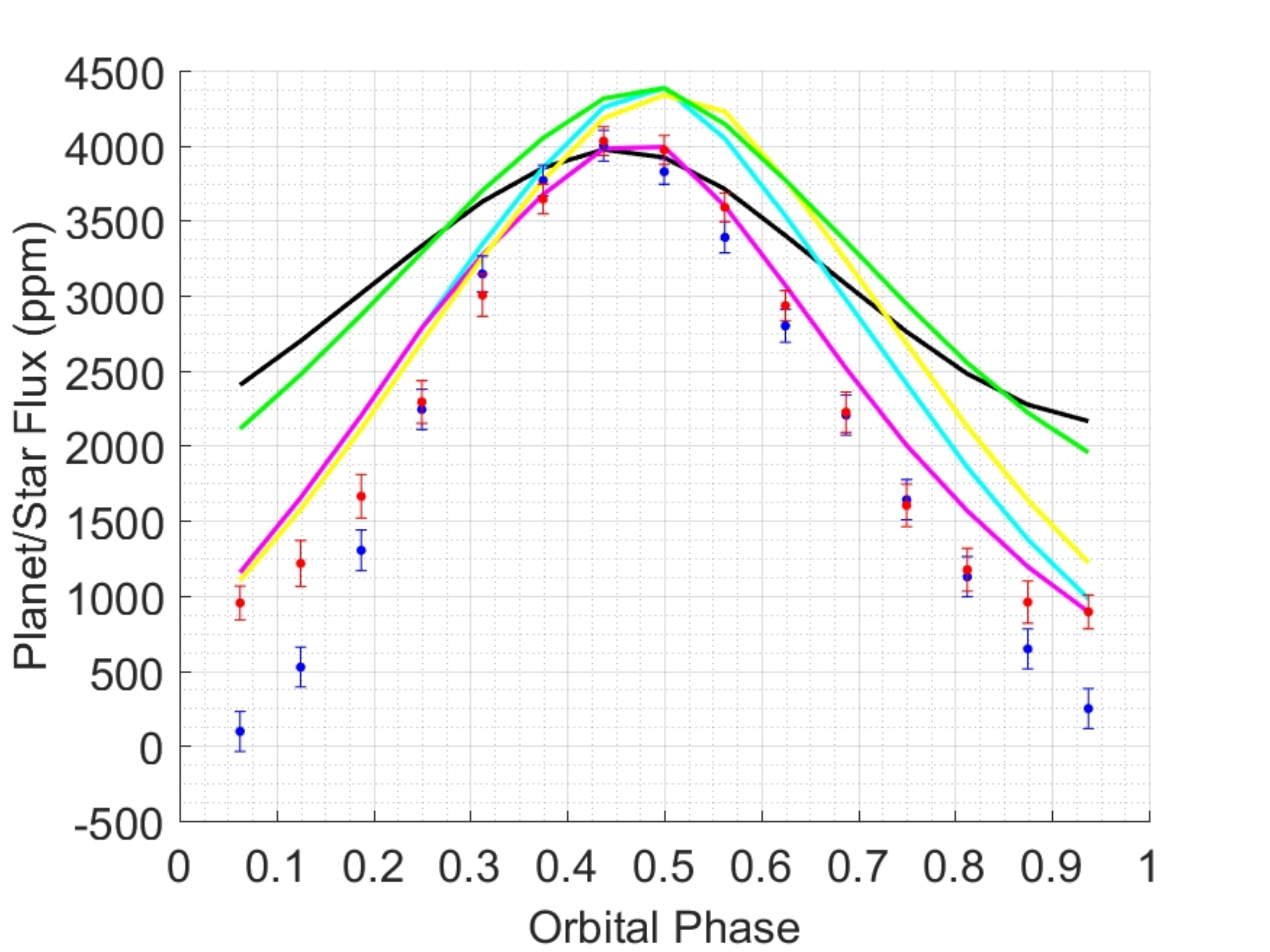}}
\caption{Phase curves for different wavelengths (panels $a$ to $i$). The planet transit happens at orbital phase 0.0 and the secondary eclipse at 0.5. The blue points are \textit{WFC3} data from \cite{2014Stevenson} and \textit{Spitzer} data from \cite{2017Stevenson}. The red points are from our re-analysis of the \textit{Spitzer} data. The colors of the curves represent the same atmospheric scenarios as in Figure \ref{fig:planet_spec}.}
\label{fig:phase-curves}
\end{centering}
\end{figure*}

The simulated and measured multi-wavelength phase curves in Figure \ref{fig:phase-curves} emphasize different properties.  First, none of the simulated phase curves from the cloudfree GCM match the data.  The simulated phase curves are somewhat flat at \textit{WFC3} wavelengths, but this is a consequence of the mean opacities we have chosen, which render the dayside-nightside contrast at $\gtrsim 1$ bar to be low.  Physically, starlight has been mostly attenuated by these pressures.  At these depths in our GCMs, the zonal winds effectively establish near-uniformity in temperature across longitude.  By contrast, the different treatment of opacities by \cite{2015Kataria} results in a different thermal structure in the deep atmosphere, which leads to different predictions for the cloudfree phase curves. At all wavelengths except 4.5 $\mu$m, the various cloudy models match the data well.  At 3.6 $\mu$m, our re-analysed phase curve is more naturally in agreement with the models.  At 4.5 $\mu$m, the model with enhanced CO$_2$ is favored.

\section{Conclusions and future prospects}
\label{sec:conclu}

We have revisited the phase curves of WASP-43b---both observationally and theoretically.  Our re-analysis of the \textit{Spitzer} data results in higher fluxes emanating from the nightside of WASP-43b, which brings the nightside spectrum into closer agreement with GCM predictions.  Confronting our GCM predictions with data leads us to conclude that clouds on the nightside of WASP-43b are necessary, and that constraints on the cloud-top pressure and longitudinal position of the cloud deck are obtained.  There is a hint that enhanced CO$_2$ abundances are present at and near the nightside of WASP-43b.

A lesson learned is that multi-phase emission spectra and multi-wavelength phase curves constrain different properties of the atmosphere.  The former are useful for constraining cloud properties, while the latter tell us if clouds are needed at all.  Phase curves at specific wavelengths inform us if specific molecules are present at enhanced abundances---in our case, CO$_2$ at 4.5 $\mu$m.

There are ample prospects for future work.  \texttt{THOR} may be upgraded to fully and self-consistently incorporate multi-wavelength radiative transfer, without the need for post-processing.  With the \textit{James Webb Space Telescope} due to launch in 2019, multi-phase emission spectra at higher spectral resolution may be obtained.  A visible, reflected-light phase curve of WASP-43b would be especially constraining, as it would directly constrain the longitudinal distribution of clouds.  In this regard, a resounding example has already been provided by Kepler-7b \citep{2013Demory}.

\section*{Acknowledgments}
J.M.M., M.M., B.-O.D. and K.H. thank the Center for Space and Habitability (CSH), Swiss National Science Foundation, Swiss-based MERAC Foundation and the Space Research and Planetary Sciences Division (WP) of the University of Bern for financial, secretarial and logistical support. B.-O.D. acknowledges support from the Swiss National Science Foundation in the form of a Swiss National Science Foundation Professorship (PP00P2-163967). We thank Daniel Kitzmann and Shang-Min Tsai for useful discussions on cloud physics and chemistry. We also thank Sandra Raimundo for instructive conversations. 

\appendix
\section{``Double grey'' radiative transfer}
\label{apxd:rad_tr}

Our simple radiation scheme in the GCM is based on the solution of the radiative transfer equation in parallel-plane layers with no scattering. The method used to represent the incoming stellar radiation solves the Lambert law equation:
\begin{equation}
\label{eq_dn_sw}
F_{sw}^{\downarrow} = (1-A)F_\star \exp{\Big(-\frac{\tau_{stellar}}{\mu_{zth}}\Big)},
\end{equation}
where $F_{sw}^{\downarrow}$ is the incoming downward stellar flux, $A$ is the planet bond albedo, $F_\star $ is the stellar constant, $\mu_{zth}$ is the cosine of the zenith angle, and $\tau_{stellar}$ is the optical depth for the stellar light. We include an extra layer above the model's domain to avoid overheating in the uppermost layer during the numerical simulations. 

The zenith angle is the angle between the zenith point and the centre of the star's disc, and the amount of incoming stellar flux at the top of the model's domain is weighted by the cosine of the zenith angle. The positive values of the cosine are related to the day-side of the planet, negative to the night-side and zero indicates the terminator. In our work we correct the solar path-length to take into account the effect of the atmospheric spherical curvature. The effective solar path-length is calculated defining the cosine of the zenith angle ($\mu$) as (\citealt{2006Li}):
\begin{equation}
\frac{1}{\mu(z)}=\frac{1}{\sqrt{1-\big(\frac{R}{R+z}\big)^2(1-\mu_0^2)}}
\end{equation}
In the equation $z$ represents altitude, $R$ the radius and $\mu_0$ the cosine of the zenith angle without the geometrical correction. In order to simplify this equation we have neglected the refraction effects. 

In the thermal radiation part, we solve for each layer the following thermal emission equation: 
\begin{equation}
\label{eq:E}
dE(\mu) = B(\tau ')\exp{\big(-\tau '/\mu\big)}\frac{d\tau '}{\mu}.
\end{equation}
The variable  $B(\tau')$ is the spectrally integrated Planck function and $\mu$ is the cosine of the emission angle relative to the normal of the layer basis. $\mu$ is always define positive where layer basis is relative to the direction of radiation. Inside the model layers we assume that the source function varies linearly with optical depth. The notation used here is the same as in \cite{2015Mendonca}:
\begin{equation}
B(\tau ') = B(T_T) + \frac{\tau'}{\tau}[B(T_B) - B(T_T)].
\end{equation}
$B(T_T)$ and $B(T_B)$ are the Planck functions for temperatures at the top and the bottom of the atmospheric layers respectively, and $\tau$ is the total optical depth of the layer. The integrated flux from the Planck function is calculated using the Stefan-Boltzmann equation.

Upon integrating Eq. \ref{eq:E} for the entire layer we can obtain two solutions for the emitted thermal radiation: upward and downward directions. The solution in the upward direction is:
\begin{equation}
E(\mu) =  B(T_T)-B(T_B) + [B(T_B) + \frac{\mu}{\tau}[B(T_B) - B(T_T)](1- \exp{\Big(-\frac{\tau}{\mu}\Big)}).
\end{equation}
The downward solution is represented by:
\begin{equation}
E^{\star}(\mu) =  B(T_B)-B(T_T) + [B(T_B) - \frac{\mu}{\tau}[B(T_B) - B(T_T)](1- \exp{\Big(-\frac{\tau}{\mu}\Big)}).
\end{equation}
As pointed out in \cite{1991Lacis} the two solutions above have to be fixed for the case of small optical depths due to the singularity in the equations. It is suggested in \cite{1991Lacis} that in the case for small optical depths these two equations can be replaced by:
\begin{equation}
E(\mu)=\sum^{n_f}_{n=1}(-1)^{n+1}\frac{B(T_T)+ nB(T_B)}{(n+1)!}(\frac{\tau}{\mu})^n
\end{equation}
for the upward direction and in the opposite direction:
\begin{equation}
E^{\star}(\mu)=\sum^{n_f}_{n=1}(-1)^{n+1}\frac{B(T_B)+nB(T_T)}{(n+1)!}(\frac{\tau}{\mu})^n.
\end{equation}
In our model we set $n_f$ to be equal to five.

The solutions for each layer are then combined in a stacked layer atmosphere configuration. The net upward thermal intensities are calculated from:
\begin{eqnarray}
\label{eq:U}
\nonumber
   U_0(\mu) &=& B(T_{int}) \\
   U_1(\mu) &=& E_1(\mu) + U_0(\mu)\exp{\big(-\tau_1/\mu\big)} \\
\nonumber 
   ... \\
\nonumber
 U_n(\mu) &=& E_n(\mu) + U_{n-1}(\mu)\exp{\big(-\tau_n/\mu\big)}
\end{eqnarray}
where $ B(T_{int})$ represents the flux coming from the planet's interior and the indeces $n$ are the layers interfaces' indeces (0 represents the lowest model interface and N the top of the model domain). The downward component is calculated from:
\begin{eqnarray}
\label{eq:D}
\nonumber
D_{N}(\mu) &=& 0 \\
D_{N-1}(\mu) &=& E_N^{\star}(\mu) +  D_{N}(\mu)\exp{\big(-\tau_N/\mu\big)} \\
\nonumber 
   ... \\
\nonumber
 D_{n}(\mu) &=& E_{n+1}^{\star}(\mu) +  D_{n+1}(\mu)\exp{\big(-\tau_{n+1}/\mu\big)}.
\end{eqnarray}
At the top we assume that the downward flux is zero. The net fluxes are obtained integrating the equations over $\mu$. In our scheme  the angular integration is calculated using a three-point Gaussian quadrature. Using this integration we have a simple and accurate method to estimate the angular integration and it is more flexible than other approximate techniques such as the diffusivity factor.

In order to improve the accuracy of the thermal source functions of each layer, we divided the model layers into two equally optically thick parts and computed the intensities separately (\citealt{2015Mendonca} and \citealt{2017Malik}). This method improves the accuracy in the thermal emission calculation in the middle of the layers and avoids the formation of local spurious peaks in the temperature profile. The solar and thermal heating/cooling rates ($\frac{dT}{dt}$) are calculated from the total flux difference across each layer:
\begin{equation}
\frac{dT}{dt}=\frac{1}{\rho C_p}\frac{dF^{net}}{dz}.
\end{equation}
In this equation $\rho$ is the atmospheric density, $C_p$ is the specific heat capacity at constant pressure and $F^{net}$ is the spectral-integrated net radiative flux.


\end{document}